\documentclass[pdflatex,sn-standardnature]{sn-jnl}% Standard Nature Portfolio Reference Style
\usepackage{lmodern}
\usepackage{subcaption}
\usepackage{fix-cm}
\usepackage{anyfontsize}
\usepackage[export]{adjustbox}  % For adding borders to images
\jyear{2025}%

\theoremstyle{thmstyleone}%
\theoremstyle{thmstyletwo}%

\theoremstyle{thmstylethree}%

\begin{document}

\title[The Unified Evaluation App for DNA Data Storage Codecs]{The Unified Evaluation App for DNA Data Storage Codecs}

\author*[1,2]{\fnm{Aleksandar} \sur{Anžel}}\email{anzela@rki.de}
\equalcont{These authors contributed equally to this work.}

\author[1,2]{\fnm{Chisom} \sur{Anyabolu}}%\email{anyaboluc@rki.de}
\equalcont{These authors contributed equally to this work.}

\author[2]{\fnm{Leon} \sur{Wimbes}}%\email{wimbes@students.uni-marburg.de, leon.wimbes@web.de}
\author[2]{\fnm{Luca} \sur{Staus}}%\email{luca.staus@uni-jena.de}
\author[2]{\fnm{Ihsan} \sur{Tri Heldian}}%\email{heldian.ihsan@gmail.com}
\author[2]{\fnm{David} \sur{Sonnabend}}%\email{Sonnaben@students.uni-marburg.de}
\author[2]{\fnm{Khawla} \sur{Elhadri}}%\email{elhadrik@staff.uni-marburg.de}
\author[2]{\fnm{Samuel} \sur{Becker}}%\email{beckersr@students.uni-marburg.de}
\author[2]{\fnm{Felix} \sur{Klein}}%\email{Kleinfe5@students.uni-marburg.de}
\author[2]{\fnm{Raffael} \sur{Schoen}}%\email{raffaelschon@rocketmail.com}

\author[2]{\fnm{Michael} \sur{Schwarz}}%\email{peter.schwarz@uni-marburg.de}
\author[3]{\fnm{Marius} \sur{Welzel}}%\email{marius.welzel@uni-muenster.de}

\author[2]{\fnm{Bernd} \sur{Freisleben}}
%\email{bernd@uni-marburg.de}

\author[3]{\fnm{Dominik} \sur{Heider}}%\email{dominik.heider@uni-muenster.de}

\author[1,4]{\fnm{Georges} \sur{Hattab}}%\email{hattabg@rki.de}

\affil[1]{\orgdiv{Center for Artificial Intelligence in Public Health Research (ZKI-PH)}, \orgname{Robert Koch Institute}, \orgaddress{\street{Nordufer 20}, \city{Berlin}, \postcode{13353}, \country{Germany}}}

\affil[2]{\orgdiv{Department of Mathematics \& Computer Science}, \orgname{University of Marburg}, \orgaddress{\street{Hans-Meerwein-Straße 6}, \city{Marburg}, \postcode{D-35032}, \country{Germany}}}

\affil[3]{\orgdiv{Institute of Medical Informatics}, \orgname{University of Münster}, \orgaddress{\street{Albert-Schweitzer-Campus 1, Gebäude A11}, \city{Münster}, \postcode{48149}, \country{Germany}}}

\affil[4]{\orgdiv{Department of Mathematics \& Computer Science}, \orgname{Freie Universität}, \orgaddress{\street{Arnimallee 14}, \city{Berlin}, \postcode{14195}, \country{Germany}}}

\abstract{
\textbf{Background:} Deoxyribonucleic acid (DNA) data storage is a paradigm with great potential for ultra-dense and durable information preservation. However, the rapid proliferation of coding schemes, or codecs, each with their own design constraints and reporting practices, has led to a fragmented landscape that lacks a standardized comparative assessment.

\textbf{Methods:} We developed an open-source, modular benchmarking platform that systematically integrates and evaluates state-of-the-art DNA storage encoding and decoding methods (codecs). Our approach uses a curated, diverse set of baseline data and applies multidimensional assessment criteria that are aligned with the consensus standard of the DNA Data Storage Alliance. These criteria include encoding/decoding throughput, computational efficiency, error correction performance across substitutions, insertions, and deletions, and cost efficiency.

\textbf{Results:} The developed platform integrates standardized wrapper functions for encoding and decoding, allows for the integration of new methods, and automates reproducible evaluations with comprehensive visual and tabular reporting. Benchmarking both contemporary and classical codecs using their default parameters and multiple metrics demonstrates that no single algorithm is optimal across all evaluated dimensions. The trade-offs between information density, success rate, runtime, and cost are quantified and shown to be critical factors in the design of future-proof formats.

\textbf{Conclusions:} Our work establishes a rigorously standardized, open-source evaluation framework that enables reproducible benchmarking, supports evidence-based codec selection, and provides the necessary foundation for translating DNA data storage from experimental research into deployable archival systems.
}

\keywords{
    DNA data storage, Coding schemes, Codecs, Benchmarking, Comparative analysis, Performance evaluation, Unified framework, Multi-algorithm comparison, Error correction, Information density, Digital data archiving}



\maketitle

\section{Introduction}\label{sec-introduction}

% ==========================================
% All in present tense.

% Describe the problem

% Describe the strategy to solve that problem

% Describe what is known and what is unknown while solving the problem

% Describe the hypothesis
% ==========================================
In the rapidly evolving field of data storage, deoxyribonucleic acid (DNA) data storage has emerged as a promising solution to meet the global demand for information preservation. 
%For 2025, the global datasphere, that is the volume of data created, captured, copied and consumed worldwide, is estimated to reach the order of $10^{14}$ gigabytes (about $1.8 \times 10^{14}$ GB, or roughly 181 zettabytes), a figure that far surpasses the capabilities of conventional storage technologies~\cite{intro-rydning2018digitization,intro-Shetty2024From, intro-statista2022DataVolume}.
%By 2025, the world's data storage requirements are projected to reach $1.75 \times 10^{14}$ GB, a figure that far surpasses the capabilities of conventional storage technologies~\cite{intro-rydning2018digitization,intro-Shetty2024From}. 
By 2030, the amount of data generated globally each year is projected to reach approximately 1 yottabyte ($10^{24}$ bytes $\approx 10^{15}$ GB). To accommodate this growth, global general storage capacity is expected to increase roughly tenfold relative to 2020, reaching approximately $3.7 \times 10^{13}$ GB. This surge is driven primarily by the exponential growth of artificial intelligence-related data, which is projected to constitute over 60\% of enterprise data access and processing by that time~\cite{huawei2024datastorage}. Due to its unparalleled storage density, long shelf-life, potential low maintenance cost, and chemical modification capabilities, DNA offers a viable alternative for the long-term archival of large amounts of data~\cite{intro-the-visual-story, intro-zettabyte}. In addition, DNA is omnipresent and biologically significant. Consequently, as long as life exists, there will be a need to understand and manipulate DNA, ensuring its resilience and preventing it from becoming obsolete.

\begin{figure}
\includegraphics[width=\textwidth]{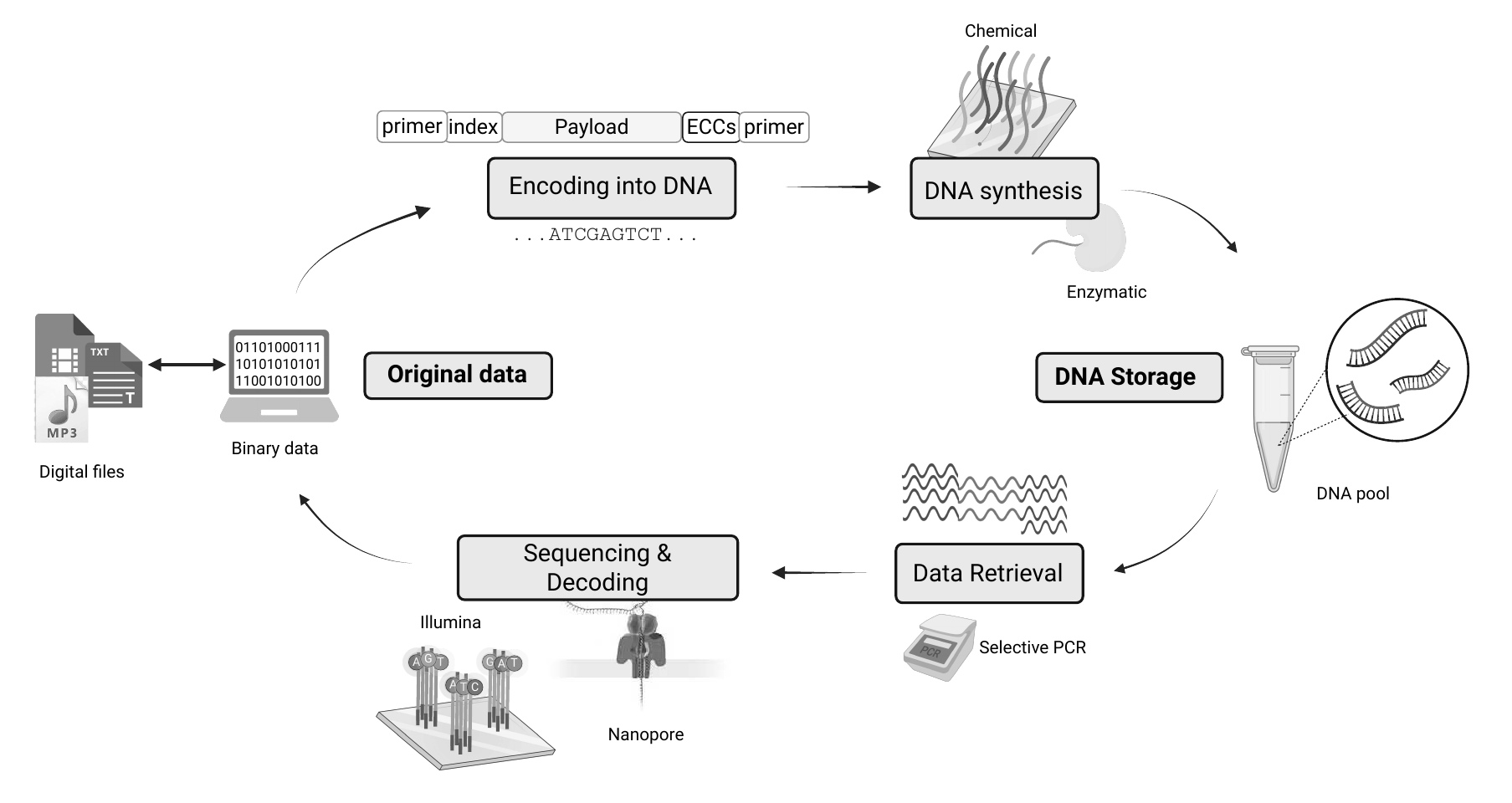}
\caption{\textbf{Overview of the processes involved in a DNA data storage system.} The DNA data storage process involves reading digital files in their binary format, encoding them into DNA sequences, synthesizing the DNA, storing it in a single- or double-stranded form, sequencing the DNA, and decoding it to retrieve the original binary data. Created in Biorender \url{https://www.biorender.com/}.}
\label{fig:process}
\end{figure}

The DNA molecule is made up of four fundamental building blocks of nucleic acids called nucleotides: adenine (A), cytosine (C), guanine (G), and thymine (T). A single strand of DNA, often called an oligonucleotide, represents an ordered sequence of these nucleotides. 
In effect, it forms a string through the alphabet~---~\{A, C, G, T\}. 
The ability to chemically synthesize any conceivable sequence of nucleotides underpins the potential to store digital information within DNA strands. By encoding binary data in nucleotide sequences, DNA can be used as a high-density storage medium, providing a novel approach to data storage. In this work, we focus specifically on \textit{in vitro} DNA-based storage systems, rather than \textit{in vivo} methods that involve storing data within living organisms.
As shown in~\autoref{fig:process}, the initial step in the \textit{in vitro}-based storage process is the conversion of digital files into binary data, which is then encoded into DNA sequences. Additional sequences, including an index, error correction codes, and primers for DNA amplification, are appended to or integrated within the oligonucleotides (oligos). The synthesis of oligonucleotides is typically achieved through chemical or enzymatic processes, resulting in the production of single-stranded DNA. This synthetic DNA is frequently stored in its single-stranded form; alternatively, a complementary strand is synthesized enzymatically, producing double-stranded DNA for storage. The selective extraction or amplification of specific oligos from the pool is achieved through polymerase chain reaction (PCR). Once extracted, the oligos are sequenced using a DNA sequencer, and the obtained nucleotide sequences are subsequently decoded to retrieve the original binary data.

Despite its potential, DNA data storage faces significant challenges that impede its widespread adoption~\cite{intro-zettabyte}. Current methods of writing (DNA synthesis) and reading (DNA sequencing) are notably slower than conventional storage technologies.
Furthermore, unlike with sequencing, synthesis costs remain too high for widespread and scalable adoption. Another critical issue is DNA preservation, which requires ensuring the long-term stability and integrity of the encoded information. The use of DNA as a storage medium is limited by appropriate temperature, humidity, pH, and radiation protection, among other factors, which makes this storage solution significantly more complicated than its competitors.
Moreover, the processes of DNA synthesis, storage, and sequencing are error-prone, which further complicates the situation. Due to the nature of DNA, these errors fall into one of three categories: insertion, deletion, or substitution of nucleotides (DNA bases). DNA degradation is closely tied to the chemical nature of the four nucleotides. For instance, when the GC content is lower than 40\% or higher than 60\%, the probability of synthesis and sequencing errors increases. This motivates encoding digital data for long-term DNA archival in a way that satisfies various conditions and constraints, thereby optimizing the DNA data storage medium.

As technological hurdles in DNA-based digital data storage are gradually overcome, the field is reaching a pivotal point. Advances in computational power, reduced costs, and accelerated DNA synthesis are establishing DNA storage as a viable alternative to traditional storage media~\cite{synthesis-better}.
Several researchers have demonstrated the feasibility of storing and successfully retrieving digital information using DNA molecules. Early works by Church \textit{et al.}~\cite{intro-church} and Goldman \textit{et al.}~\cite{intro-Goldman2013} were pivotal in bringing DNA data storage to the limelight.
With advancements in DNA synthesis and sequencing technology, the end-to-end workflow for DNA-based data storage has been well established~\cite{intro-takahashi2019demonstration}. However, current technology still imposes specific constraints on DNA sequences, such as GC content and the length of homopolymers (a sequence of identical nucleotides repeated multiple times). Addressing these constraints while achieving high storage density is vital for the efficient implementation of DNA data storage systems.
Based on these constraints, newer methodologies for using DNA as a storage medium continue to emerge, pushing DNA data storage toward practical applications. 
These methodologies promise not only larger storage capacity and longer lifespan but also address various information and coding theoretic aspects, such as storage channel capacity and the design of error correction codes~(ECC) tailored to specific errors encountered in DNA data storage. ECC development efforts have also targeted constrained coding~\cite{intro-lochel2021fractal, intro-yin2021design, intro-limbachiya2018family}, random access in DNA data storage systems~\cite{intro-TabatabaeiYazdi2015,intro-Organick2018, Cao2025}, and image storage solutions~\cite{intro-Ping2022}, enhancing the robustness and versatility of DNA data storage systems.

Despite the rapid development of DNA coding schemes (codecs) and researchers' dedicated efforts to address the challenges of DNA data storage, a fair and unified comparison of these methods is lacking. 
The only notable exceptions are the works of Ping \textit{et al.}~\cite{chamaeleo-Ping2020.01.02.892588} and Gimpel \textit{et al.}~\cite{grass-comparison}, the former introducing a modular Python library implementing five DNA codecs, and the latter providing a wet-lab evaluation of six DNA codecs.
Nevertheless, these works are still limited by different factors.
On the one hand, the study by Ping \textit{et al.}~\cite{chamaeleo-Ping2020.01.02.892588} evaluates only the encoding and decoding runtimes of each method, without considering additional performance metrics (\textit{e.g.}, cost, size, \textit{etc.}). Moreover, the library supports only limited error simulation: substitutions are modeled, while insertions and deletions are not. 
On the other hand, Gimpel \textit{et al.}'s benchmarking approach~\cite{grass-comparison} is constrained by a fixed selection of codecs, workflows, and harmonized parameters, limiting generalizability as new algorithms and experimental techniques emerge, while also making it difficult for the community to expand or customize benchmarks as field requirements change. Additionally, their solution does not consider specialized sequence constraints or novel performance metrics, provides limited reproducibility and transparency for users lacking significant technical expertise, and does not include visual analytics or an accessible dashboard interface for rapid inspection and interpretation of benchmarking results.
Beyond these studies, newly developed DNA codecs are usually evaluated using a limited set of metrics and compared with a selection of current state-of-the-art approaches.
This practice is common in nearly all previously published works in this field, and it persists in even the most recent methods such as DNA-Aeon~\cite{methods-dnaaeon-Welzel2023} and DNAformer~\cite{new-method-daniella-2025}.

The diverse landscape of these codecs, which spans various programming languages, software architectures, and a plethora of parameters of individual codecs, poses challenges for new and existing researchers. These researchers seek to accurately assess the current research landscape and effectively evaluate their newly developed methods. Furthermore, many existing codecs have suboptimal software architecture. This hinders effective program-based validation and impedes efforts to improve these methods. The absence of well-designed, modular architectures complicates enhancing individual methods and obstructs developing standardized benchmarking tools for the field as a whole.

As outlined by the DNA Data Storage Alliance, the establishment of a unifying evaluation framework is of paramount importance~\cite{SNIA2025}. The Alliance's work emphasizes the necessity for standardized methodologies to systematically assess and integrate DNA codecs. Such standardization enables rigorous comparison across critical parameters including storage density, cost-efficiency, write and read throughput, long-term data stability, encoding and decoding performance, scalability, interoperability with extant technologies, and adaptability to diverse data formats. Adherence to these standardized evaluation criteria is instrumental in advancing the field by fostering consistency, reproducibility, and the effective translation of research innovations into practical DNA data storage solutions. %energy efficiency, environmental impact, 
%Moreover, existing encoding methods perform variably on different data structures or application scenarios, and establishing a systematically, integrated evaluation platform to flexibly provide corresponding storage solutions based on different storage needs will promote the adoption of more DNA data storage applications.
%Therefore, it is necessary to consolidate well-established encoding methods within a robust software structure and provide flexible usage for research, application, and optimization. 
%This research aims to develop a comprehensive evaluation framework and a unified platform for DNA data storage encoding methods.

%conclusion
This endeavor seeks to establish a modular, open-source, and standardized user-friendly web app that facilitates an objective comparison of various codecs. The primary objective is to guide further advancements in the field. The proposed app identifies the most promising DNA data storage solutions by addressing known challenges and incorporating various evaluation criteria, promoting their practical implementation. The unified app makes three key contributions to the field: a comprehensive library of existing codecs; a robust benchmarking system that evaluates integrated codecs using various criteria; and an interactive testing module that allows users to simulate using DNA as a data storage medium with custom files and a selection of integrated codecs. Our main goal is to speed up innovation in DNA data storage by working together openly and with the community. By encouraging contributions from researchers and developers, we aim to cultivate a dynamic ecosystem that evolves alongside the field and promotes standardization and reproducibility in DNA data storage research.

%Relevant PDF~\url{https://dnastoragealliance.org/wp-content/uploads/2021/06/DNA-Data-Storage-Alliance-An-Introduction-to-DNA-Data-Storage.pdf}.

\section{Methods}\label{sec-methods}
%talk about its exclusiveness to python
%intial data set - 5 textfiles, 2 images, 3 audio files
%The DNA simulation using MESA - To our knowledge, MESA is the only currently available simulator able to model the full DNA data storage channel, including synthesis, PCR amplification, storage and sequencing. It uses either published error profiles or user-defined rates. The sequencing module allows to simulate reads from different sequencing platforms such as Illumina, PacBio and Nanopore. In addition, this tool offers the possibility of assessing the quality of the DNA fragments regarding their content (i.e. whether the biological constraints are respected or not) 
%Two wrapper functions- encode(file\_path, bool array) which returns list of DNA strings and decode(file\_path, list od DNA strings) which returns a NumPy bool array

Our work proposes a unified evaluation app for DNA data storage codecs. The app is primarily designed to evaluate the characteristics of different encoding/decoding strategies based on the same baseline data set.
To facilitate evaluation and user access, we transformed the original idea into a web application. The benchmark page and file upload page are two integral parts of the web application and core functionalities of the unified app. The former implements the evaluation of integrated codecs using a baseline data set. The latter enables users to evaluate integrated codecs using a file they uploaded, with the added option of simulating the long-term feasibility of storing DNA data.

% This structured approach ensures a comprehensive evaluation of different encoding/decoding strategies, facilitating the comparison of performance measures such as encoding and decoding time, file sizes, synthesis cost, and data integrity across various DNA data storage encoding methods.
This section provides an in-depth look at the web app's user interface, workflow, integrated codecs, evaluation system, and file upload functionality.

\subsection{User Interface and Workflow}\label{sec-user-interface-and-workflow}

The unified application is presented through a web interface that is interactive, dynamic, and user-friendly. This interface is designed to guide users seamlessly through the stages of using DNA as a data storage medium. The interface consists of five pages: Home, Benchmark, File Upload, About, and Contact Us. The Home and Contact Us pages provide essential information about the app, including ways to contribute and sources of support. The About page contains the information about the benchmark data, evaluation metrics, and supported codecs. The Benchmark and File Upload pages encapsulate the core functionalities of the application. The following text focuses on these two functionalities and describes the user interface and workflow within these two pages.

\subsubsection{Benchmark Page}\label{sec-benchmark-page}

The benchmark page presents the evaluation results of integrated codecs defined in~\autoref{sec-encoding-methods}, using the baseline data set and evaluation criteria outlined in~\autoref{sec-evaluation-system}. The results presented on this page are calculated using the procedure outlined in~\autoref{sec-file-upload-page}. Upon accessing the page, users are presented with a summary benchmark data set given in a tabular format, containing aggregated evaluation results across multiple runs. An interactive expander element containing the detailed benchmark results in a tabular format is located immediately below the summary benchmark. The purpose of the expander element is to reduce cognitive load by omitting the larger detailed benchmark data set from immediate view. By following the well-known visualization mantra ``Overview first, [\ldots], then details-on-demand'' by Schneiderman~\cite{methods-overview-first}, users are not overwhelmed by the details but can still inspect them closely if necessary. In addition, users can interact with the tables by clicking the cells containing the column names to easily sort the tables in either ascending or descending order based on the values in that column. Below both benchmark tables are buttons that allow users to easily download them in \texttt{CSV} format.

Below the expander containing a detailed benchmark table lies the first visualization, which presents the data information density of each codec across all runs and all data. This ratio is visualized using a whisker chart to capture the variance of information density for each codec. The chart includes interactive elements such as linked zooming and tooltips, encouraging deeper exploration of the results. The whisker representation mirrors the volatility plots proposed in the DNA Data Storage Alliance's white paper for throughput benchmarking, whereby inter-quartile ranges reveal violations of the ``encode faster than synthesis'' heuristic~\cite{SNIA2025}.

The final section of the benchmark page provides a multi-select element, a chart type selector, and two small-multiple charts~---~one visualizing the runtime and the other visualizing the costs of each codec during both the encoding and decoding process across the selected benchmark files. The multi-select element is by default loaded with all files that are part of the benchmark data set; however, users can selectively analyze only some of the files using this widget. Immediately below the multi-select element lies a chart type selector that allows users to choose between whisker and bar charts. The selection is then reflected in both small-multiple charts below.

\begin{figure}
    \centering
    \includegraphics[width=0.98\textwidth, frame]{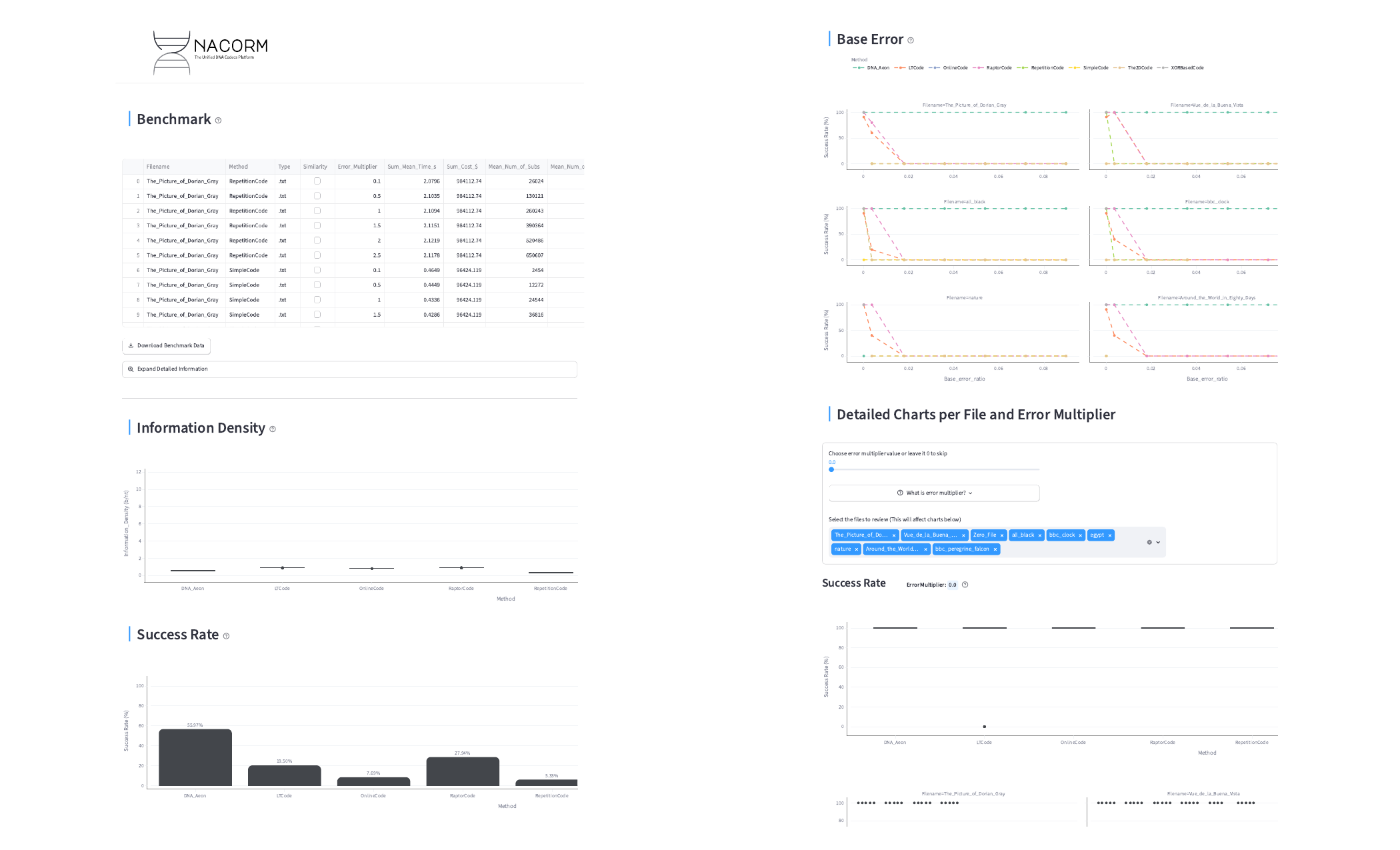}
    \caption{\textbf{Overview of the Benchmark page.} The benchmark page contains performance evaluation of various DNA codecs. The results are presented in both tabular and graphical formats, allowing users to compare encoding efficiency, speed, and other relevant metrics comprehensively.}
    \label{fig: benchmark_page}
\end{figure}

Each version of a small-multiple chart depicts a different file from the benchmark data set. Within each small multiple, codecs are placed on the x-axis, while either time or cost is placed on the y-axis. For each codec, colors of whiskers or bars indicate either an encoding or decoding process. Additionally, users can easily download all chart in either vector or raster image formats. The benchmark page, with all elements mentioned earlier, is shown in \autoref{fig: benchmark_page}.

\subsubsection{File Upload Page}\label{sec-file-upload-page}

\begin{figure}[htb]
    \centering
    \includegraphics[width=\textwidth, frame]{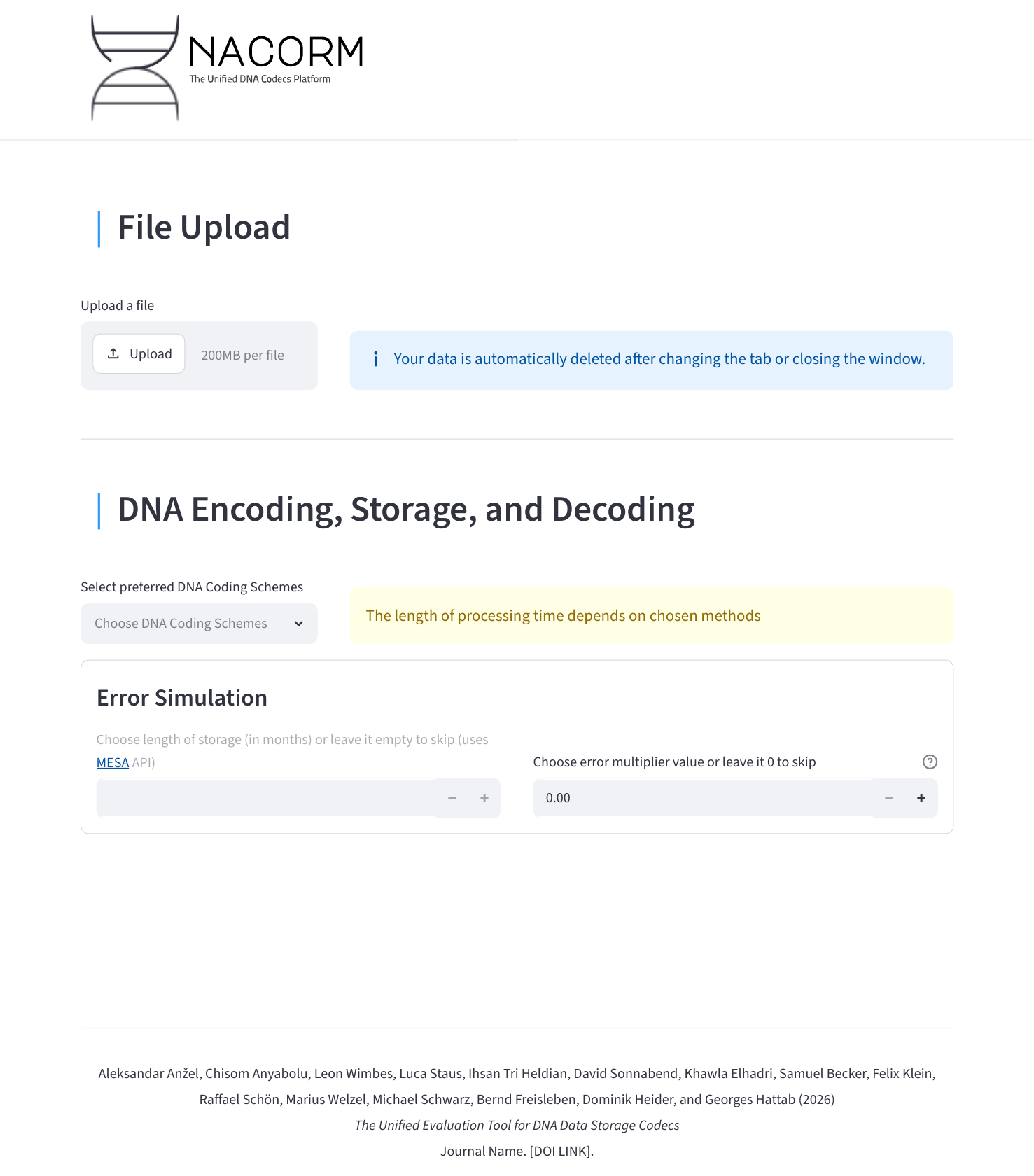}
    \caption{\textbf{Overview of the Upload page.} Screenshot of the upload page where users can submit their own files to benchmark across multiple DNA codecs. The interface supports simulation of DNA degradation using MESA~\cite{methods-mesa}, enabling users to observe the impact on coding performance and error resilience with real-time benchmark results.}
    \label{fig: upload_page}
\end{figure}

\begin{figure}[htbp]
    \centering
    \includegraphics[scale=0.55]{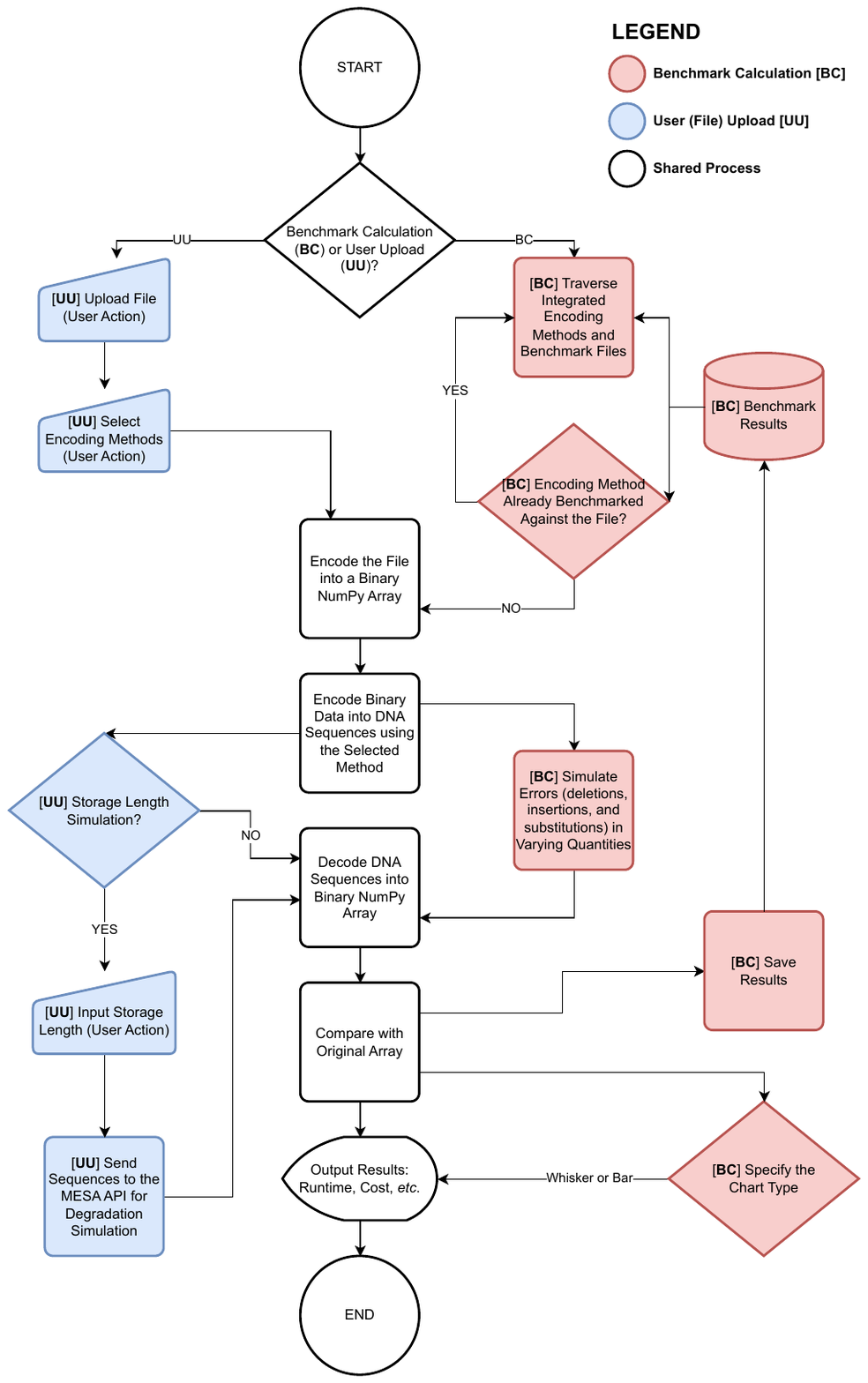}
    \caption{\textbf{Flowchart depicting the workflow of our web app.} The chart outlines the encoding, storage, and decoding processes. Key user interactions and decision points are highlighted to enhance understanding of the methodology.}
    \label{fig: flowchart}
\end{figure}

While the benchmark page showcases precalculated evaluation results derived from the workflow accessible on the file upload page, the file upload page enables users to interactively explore the processes of DNA data encoding, storage, and decoding in real-time. Therefore, we structure the current section into three distinct parts, each addressing a different segment of the pipeline.
The pipeline encompassing both benchmark and file upload functionalities is depicted in~\autoref{fig: flowchart} and given as a supplementary file. The file upload page, with all elements described in subsequent sections, can be seen in \autoref{fig: upload_page}.

\paragraph{DNA Encoding Process}\label{par-dna-encoding-process}

Upon accessing the file upload page, users are prompted to upload a file of their choice, which can be in any format. Upon upload, the file is automatically converted into a NumPy~\cite{methods-numpy} boolean array, which emulates the file's binary representation. This approach, while inherently less efficient in terms of speed and memory usage compared to utilizing a true binary format, is deliberately chosen to ensure compatibility across all codecs. By standardizing this representation, we establish a universal benchmarking framework that facilitates fair and consistent evaluation of diverse coding techniques.
Following the upload and binary transformation, users are presented with a drop-down menu that contains a list of DNA codecs.
Users can choose one or more DNA codecs integrated into the app and described in~\autoref{sec-encoding-methods}, allowing them to easily explore and evaluate different coding strategies for the uploaded file.
The selection of codecs initiates the encoding process. The binary representation of the original uploaded file (defined in a NumPy array) is parsed as input to the selected DNA codecs. The process runtime depends greatly on the size of the uploaded file and the number and type of selected DNA codecs.

As a result, each DNA codec outputs a list of DNA sequences, with the number of sequences and their structure varying significantly between codecs. Each DNA sequence is represented as a string where each character is an element of the alphabet $\{A, C, G, T\}$. Moreover, for each codec, this step in the pipeline outputs the runtime, information density, coding cost, and the size of the intermediate file that contains the list of DNA sequences.

\paragraph{DNA Data Storage}\label{par-dna-data-storage}

To better compare DNA codecs and simulate DNA data degradation during storage, we employed two different strategies. First, for the creation of the benchmark data set and each run, our platform incorporated DNA degradation using error rates observed by the HEDGES~\cite{hedges} authors with a high mutagenesis kit, with a substitution rate of 0.0238, a deletion rate of 0.0082, and an insertion rate of 0.0039 as the baseline. Starting with a multiplier of 0 (no degradation) and gradually increasing it in steps of 0.5 up to 2.5, we tested the robustness of all codecs.

The second strategy, used on the file upload page in addition to the first, relies on the DNA data storage simulator MESA~\cite{methods-mesa}. In addition to selecting the codec, users can optionally specify the DNA storage length in months. This integer value, along with the list of DNA sequences outputted by each of the selected DNA codecs, is parsed to the MESA's API~\cite{methods-mesa}. MESA enables users to simulate DNA sequence degradation due to various storage properties such as environmental temperature, synthesis and sequencing methods, storage hosts, \textit{etc.} Given the extensive range of simulation parameters available in MESA, we fixed almost all of them to their default values and provided only the storage length as a user-selectable option. This decision was made to streamline the user experience and reduce overhead, allowing users to focus on the primary factors that significantly impact DNA degradation. By limiting user input to essential parameters, we minimize the risk of errors that could arise from misconfiguration while ensuring that results remain consistent and reproducible across different analyses. Moreover, MESA is only used on the upload page because its detailed error simulation can be time-consuming, whereas the first strategy is much faster and more suitable for benchmarking purposes. This approach not only enhances usability for those who may lack deep expertise in DNA storage conditions but also facilitates more straightforward comparisons of results across different experiments.

The output of this process provides a list of DNA sequences for each selected codec. Importantly, both strategies introduces slight changes (degradation) to each sequence during simulation.
Users may choose to omit the storage length value (in the case of the second strategy) or set the degradation multiplier to 0 (in the case of the first strategy) if they do not wish to simulate DNA data storage degradation (aging, physical mishandling, suboptimal storage conditions, enzymatic activity). This allows them to focus solely on evaluating the robustness of each DNA codec against file size and file content.

\paragraph{Clustering and Multistrand Alignment Considerations}

In DNA data storage workflows, recovering the original data requires preprocessing of sequencing reads that typically involves \textit{clustering} similar strands followed by \textit{multistrand alignment} to generate a consensus sequence for decoding~\cite{SNIA2025}. Clustering groups noisy or erroneous copies of oligos, usually using Levenshtein distance or similar metrics, which is computationally expensive with complexity at least quadratic in the number of reads~\cite{SNIA2025,shomorony2022}. Multistrand alignment further refines sequences and corrects indels and substitutions through exact or heuristic algorithms whose computational costs vary from exponential to polynomial but remain substantial~\cite{SNIA2025}. Because clustering and alignment algorithms are closely intertwined with the error-correction and decoding logic, and differ substantially among codecs, our benchmarking framework does not isolate these steps as a separate, uniform preprocessing stage. Instead, each integrated codec internally implements or omits clustering and alignment according to its design principles, error model assumptions, and algorithmic strategies. This design choice preserves the authenticity of each codec's performance, respects methodological heterogeneity, and aligns with best practices for evaluation standardization in DNA storage systems~\cite{SNIA2025}.

\paragraph{DNA Decoding Process}\label{par-dna-decoding-process}

The resulting list of DNA sequences from the DNA data storage step of the pipeline is used as input for decoding by the selected DNA codecs. Similar to the encoding process, the runtime of this procedure is significantly influenced by the size of the uploaded file and the number and type of selected codecs. After each list of DNA sequences is decoded back into a boolean NumPy array by the chosen DNA codecs, a comparison with the original array is performed to ensure data integrity. The outcome of this operation is a boolean value that indicates whether the selected DNA codecs successfully retrieved the decoded information as it was originally present in the file. Additionally, similar to the encoding process, the output includes the runtime, information density, decoding cost, and size of the intermediate file that contains a boolean NumPy array for each codec.

The results collected during the encoding and decoding processes are presented in both tabular and visual formats. The tabular representation mirrors that of the benchmark page, first showcasing the summarized results and then providing detailed evaluation outcomes. Both tables can be easily downloaded in \texttt{CSV} format using the button located immediately below each table.

Below the tables, four vertically stacked charts display different results for both the encoding and decoding processes: success rate, information density, runtime, and cost. The success rate chart illustrates the similarity between the original file and the file that has undergone encoding and decoding; the information density chart shows the difference in size between the encoded file and its original counterpart; while the cost chart visualizes the associated costs for each process. Lastly, the runtime chart shows the runtime of the encoding and decoding processes.

\subsection{Coding Schemes (Codecs)}\label{sec-encoding-methods}

The study integrates eight codecs: Repetition codec~\cite{intro-Goldman2013}, Simple codec~\cite{intro-church}, 2D codec~\cite{methods-2d-code}, XOR-based codec~\cite{methods-xor-code}, DNA-Aeon codec~\cite{methods-dnaaeon-Welzel2023}, and three near-optimal rateless erasure codecs (NORECs), namely NOREC4DNA Raptor-based codec~\cite{norec4dna, methods-raptor}, NOREC4DNA Online-based codec~\cite{norec4dna, methods-online-code}, and NOREC4DNA LT-based codec~\cite{norec4dna, methods-lt-code}. 
To enhance the readability, maintainability, and extensibility of these codecs and the overall DNA data storage process, we systematically restructured or adapted all original codecs. This was achieved by following two key principles. First, if a codec was implemented in an outdated or less commonly used programming language, such as Pascal~\cite{methods-pascal} or Perl~\cite{methods-perl}, we reimplemented it in Python~\cite{methods-python}. Second, for codecs developed in modern programming languages that outperform Python in speed (\textit{e.g.}, C++~\cite{methods-cpp}), we created a wrapper script in Python to ensure uniform usage of all codecs across our app.

Since each codec can employ different parameters that directly influence its performance and error-correction efficiency, we selected default parameter settings based on the respective studies. In cases where multiple parameters were possible, we carefully chose the most representative configuration as the default to ensure meaningful comparisons across all schemes, while also selecting parameters that provide each codec with a flexible configuration capable of encoding and decoding a wide variety of files, even if this flexibility may come at the cost of reduced efficiency. Below is a list of the specific parameters of each codec we used in our platform. Additional, less relevant parameters are available in the GitHub repository.

\begin{itemize}
    \item \textbf{Repetition codec}: No parameters.
    \item \textbf{Simple codec}: Default parameters~---~a segment length of 12 bytes, with the maximum number of ``oligos'' set to $2^{19} = 524,288$.
    \item \textbf{2D codec}: Default parameters~---~\textbf{(a)} In the inner encoder, the Galois field size ($p$) is 47, the message length is 39, the payload length is 33, the symbol size is 1, and the p-factor is 2. \textbf{(b)} In the outer encoder, the Galois field size ($p$) is 47, the message length is 713, the payload length is 594, the symbol size is 30, and the p-factor is 7.
    \item \textbf{XOR-based codec}: Default parameters~---~12 nucleotides per primer, a maximum of 6 nucleotides per byte, the minimum strand length follows the equation $27 + (10 \times \texttt{MAX\_NUCLEOTIDES\_PER\_BYTE})$, the maximum payload index gap and the minimum number of larger payloads are 10, and the minimum XOR size and step are 2 and 1, respectively.
    \item \textbf{NOREC4DNA Raptor-based codec}: Overhead value is 0.9, headers are included, the error detection algorithm is Reed-Solomon, the upper bound threshold for dropping packets is 1, chunk size is 75, seed value is 2, and 7 repair symbols are used.
    \item \textbf{NOREC4DNA Online-based codec}: Same parameters as Raptor-based codec, with additional parameters: epsilon calculated per file and the quality value at 8.
    \item \textbf{NOREC4DNA LT-based codec}: Same as Raptor-based codec, but with the upper bound threshold for dropping packets changed to 1.0, 8 repair symbols, and the implicit mode set to enabled.
    \item \textbf{DNA-Aeon codec}: Default parameters~---~chunk size is 50, the error detection algorithm is ``crc'' with the header length of 8, and the package redundancy value set to 0.4.
\end{itemize}

The restructuring effort, promoted by this study, aims to create a robust and flexible framework that simplifies the integration of new codecs. Central to this framework are two key wrapper functions: \texttt{encode(file\_path, bool\_array)} and \texttt{decode(file\_path, list\_of\_dna\_strings)}, implemented for each codec.
The \texttt{encode} function accepts either the path to a file or a corresponding NumPy boolean array that represents the binary data of the file. It processes this data using the selected codec and returns a list of DNA strings that represent the encoded sequence. Each DNA string in the list corresponds to a word where each character is an element of the alphabet $\{A, C, G, T\}$. In contrast, the \texttt{decode} function takes either the path to a file or a list of DNA strings and applies the chosen decoding algorithm to reconstruct the original binary data, returning it as a NumPy boolean array.
Each encoding or decoding operation is executed as a separate process to ensure platform robustness and prevent crashes, as each codec handles inputs differently, particularly in cases where the data may contain degradation or noise.
These wrapper functions provide a flexible foundation for the system, enabling seamless integration of future codecs. By standardizing input and output formats through these functions, researchers can easily integrate new algorithms or adapt existing ones without extensive modifications to the overall codebase. This modular approach facilitates rapid development and testing of new methods while ensuring that the framework remains adaptable to the evolving landscape of DNA data storage technologies.

\subsection{Evaluation System}\label{sec-evaluation-system}

To evaluate all integrated codecs uniformly, we used a carefully curated baseline data set alongside multiple evaluation criteria.
This baseline data set serves as a reference point, providing a consistent foundation for comparing various codecs. It was meticulously selected to accurately represent the types of data the codecs will encounter in practical applications.
Building on the criteria proposed by the DNA Data Storage Alliance for DNA codecs~\cite{SNIA2025}, we structured our benchmarking suite around five orthogonal dimensions: \textbf{(1)}~scalability, \textbf{(2)}~computational performance, \textbf{(3)}~error‐detection and -correction efficacy, \textbf{(4)}~compatibility and specification transparency, and \textbf{(5)}~biochemical constraint adherence. These axes map one-to-one to the DNA Data Storage Alliance recommendations of sub-quadratic algorithmic complexity, throughput‐per-watt targets, explicit success probabilities for substitutions/insertions/deletions, Rosetta-Stone style codec format descriptions, and GC/homopolymer constraints, respectively.

In the following sections, we will describe both the baseline data set and the evaluation criteria in detail. We will outline the selection process for the baseline data set, highlighting the importance of using diverse data types to test the robustness of each codec. Additionally, we will discuss how each evaluation criterion was determined and the rationale behind its inclusion in our assessment framework.

\subsubsection{Baseline (Benchmark) Data Set}\label{sec-baseline-data-set}

\begin{figure}[htb]
    \centering
    \includegraphics[scale=0.3]{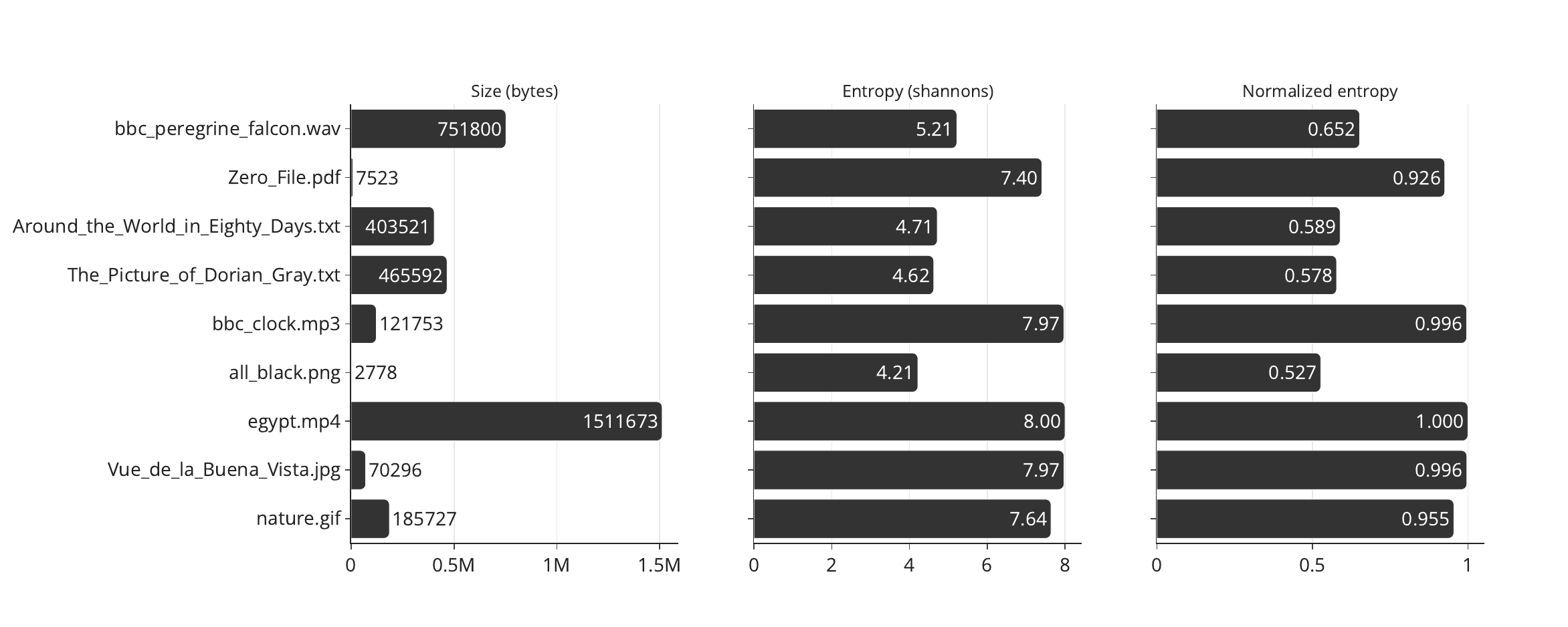}
    \caption{\textbf{Overview of benchmark data set characteristics for various file types.} Left: Distribution of file sizes in bytes across all tested formats. Middle: Measured Shannon entropy values for each file, indicating the degree of redundancy and randomness present in the data. Right: Normalized entropy values for each file, illustrating the uniformity of byte distributions.}
    \label{fig: data_set_summary}
\end{figure}

\begin{figure}[htb]
    \centering
    \includegraphics[scale=0.3]{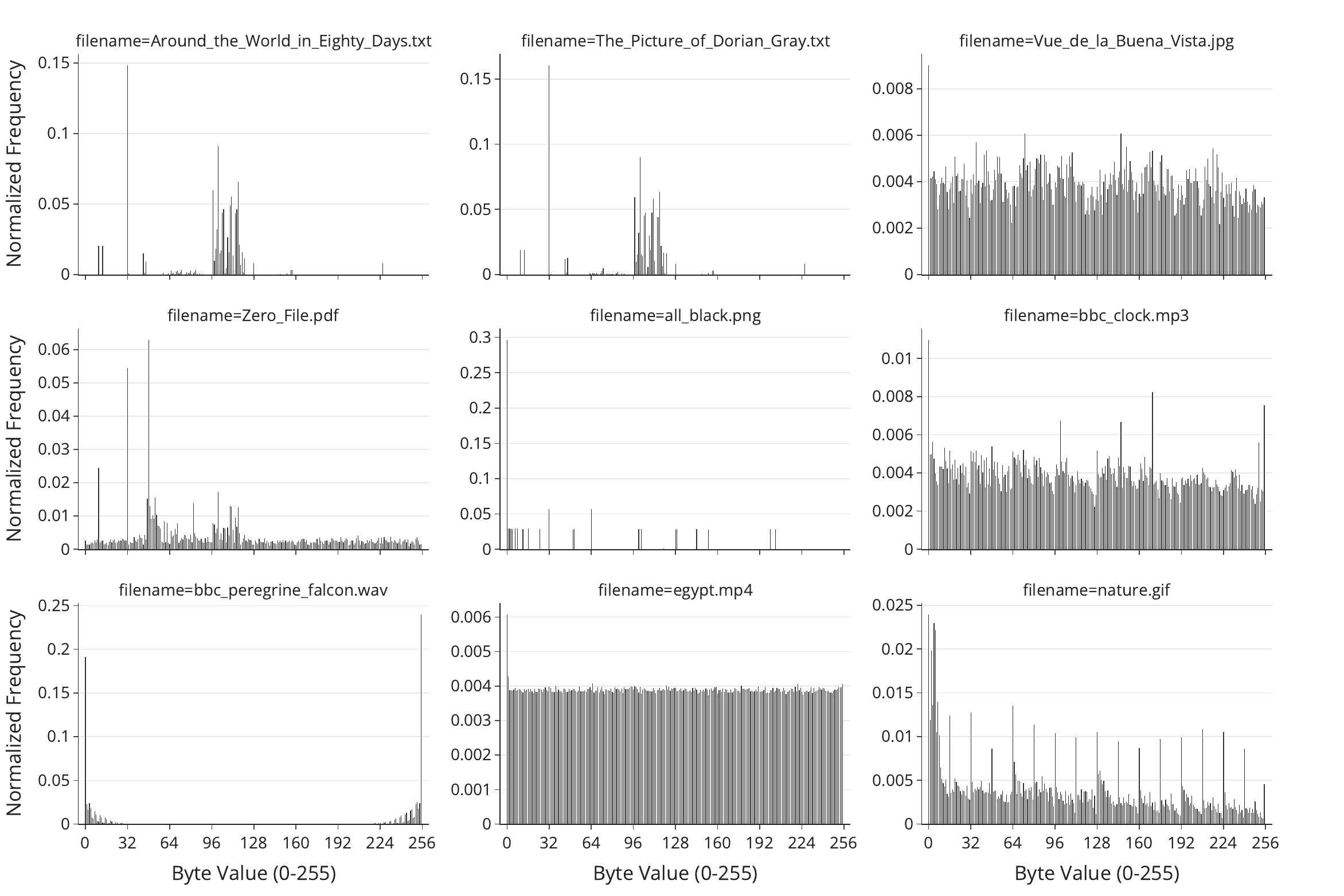}
    \caption{\textbf{Histograms displaying the byte value distributions for representative files from the benchmark data set.} Each histogram visualizes the frequency of byte values across diverse file types (PDF, TXT, WAV, MP3, MP4, PNG, GIF, JPG), revealing structural patterns, regions of redundancy, or randomness that influence algorithmic performance.}
    \label{fig: data_set_histogram}
\end{figure}

To thoroughly evaluate the performance of the DNA data storage codecs, this study utilized a diverse set of benchmark test files. The set consists of two text, three image, two audio, one video, and one document file, carefully selected to represent a broad spectrum of file types and contents commonly encountered in digital storage. This diverse data set allowed us to examine each codec on a variety of data modalities, revealing whether the method's performance relied on the underlying semantics or structure of the data, or was more influenced by the noise characteristics of the encoding (synthesis) and decoding (sequencing) processes.
Each file in the data set was encoded, underwent error simulation, and decoded five times for each codec.
This comprehensive assessment ensured a thorough evaluation of the capabilities of the codecs.
Overall sizes of the files contained within the benchmark data set can be seen in \autoref{fig: data_set_summary}. In addition, the byte value distribution of each file is presented in histograms as part of \autoref{fig: data_set_histogram}. We give a brief description of each benchmark file below.

First, the two text files that are part of the data set are books ``Around the world in eighty days'' by Jules Verne~\cite{methods-around-the-world} and ``The Picture of Dorian Gray'' by Oscar Wilde~\cite{methods-picture-of-dorian}. Inclusion of the two natural text files allowed for the efficient assessment of the codecs in handling simple, low entropy, ASCII-based data and their scalability when processing larger volumes of textual information.

Second, the three image files in standard formats were included to evaluate how well the codecs manage binary data with higher entropy, typical of visual media. Besides two images with higher entropy in \texttt{JPEG} and \texttt{GIF} formats, caused by their high compression rates, we created a lower entropy \texttt{PNG} image that contained only black pixels. The images were selected to represent different resolutions, color depths, and entropies providing insights into how effectively the methods compressed and preserved visual information. 

Third, the two audio files in \texttt{MP3}~\cite{methods-bbc-clock} and \texttt{WAV}~\cite{methods-bbc-peregrine} formats are incorporated into the benchmark tests to further challenge the codecs. These files tested the capability of the system to handle continuous, non-textual data. Due to different compression algorithms and formats the two files have completely different byte-value profiles~---~one has a uniform and flat while other has highly skewed histogram of byte values, as seen in~\autoref{fig: data_set_histogram}. The audio files presented an intricate test for maintaining fidelity in waveform-based data, which is more intricate than text or image data.

Fourth, a video file in \texttt{MP4} format was also included to assess the codecs' performance with multimedia data that combines both audio and visual components. This file represents a more intricate data type, requiring the codecs to effectively manage large volumes of information while maintaining synchronization between audio and video streams. The choice of an \texttt{MP4} file is particularly relevant, as it is the biggest in size, contains the highest Shannon entropy of 8 bits per byte, and the highest normalized entropy of 1, reflecting a uniform byte distribution across all possible values. The uniform byte distribution, as illustrated in~\autoref{fig: data_set_histogram}, is particularly significant because it can cause the presence of homopolymers following the encoding process, which can present challenges for various DNA codecs.

Finally, we incorporated a \texttt{PDF} document formatted in A4 size that contained only repeated instances of the character 0 (zero). This specific choice was made to evaluate how well the codecs handle uniform data, low in size but high in entropy, as evidenced by \autoref{fig: data_set_summary} and \autoref{fig: data_set_histogram}.

\subsubsection{Evaluation Criteria}\label{sec-evaluation-criteria}

% As with other storage media, modeling DNA as a noisy channel allows us to utilize well-established information theory to examine its properties, as well as the characteristics of corresponding encoding methods. However, unlike other storage media, DNA noise is characterized by DNA-specific errors that can occur at both intra- and inter-molecular levels. 

% On the one hand, intra-molecular errors, such as nucleotide insertions, deletions, and substitutions, typically arise during synthesis and sequencing processes. These errors are represented by the addition, removal, or exchange of one or more nucleotides from the DNA molecule, respectively.

% On the other hand, inter-molecular errors arise primarily from the processes involved in DNA storage and handling, particularly within a DNA pool. This pool is a direct consequence of the synthesis process and contains multiple copies of each DNA strand, which have been synthesized and stored in an unordered fashion. Hence, inter-molecular errors are closely linked to the characteristics of the synthesis process and can significantly affect the integrity of the DNA pool. Consequently, reading or sequencing the DNA from this pool corresponds to randomly sampling some DNA molecules~---~a process we refer to as DNA handling. Both handling and DNA degradation due to the environmental factors bring the potential loss of entire molecules from the pool.

Consistent with both DNA Data Storage Alliance codec guidance~\cite{SNIA2025} and recent comparative studies of fountain, HEDGES~\cite{hedges}, and arithmetic-based methods~\cite{new-method-daniella-2025,methods-dnaaeon-Welzel2023}, we evaluate each codec along four quantitative metrics~---~information density, success rate (data integrity), runtime, and cost.

First, \textbf{Information Density} refers to the amount of meaningful or useful information contained within a given unit of data or space. In data coding and compression contexts, information density is typically calculated as the ratio between the size of the original data and the size of the encoded or compressed data. For example, if an input file is 300 bits and it is encoded into a DNA strand of 100 nucleotides (bases), the information density is 3 (300 bits / 100 bases). This ratio indicates how efficiently the codec stores information: a higher value means more data is packed into each unit of the storage medium.
Moreover, some codecs report information density based solely on the payload, which can distort the true information density of the coded sequence. However, we argue that this metric should encompass all sources of redundancy, including both inner code overhead and the supplementary sequences introduced by outer error correction codes. Therefore, our approach computes information density using the entire binary or DNA sequence, explicitly including headers and any redundant data.

Second, \textbf{Success Rate} is evaluated by comparing the original binary data with the information retrieved after decoding the DNA sequences. This metric reflects the robustness of each codec against DNA degradation. When assessed across different base error ratios, it illustrates the codec's resilience to varying levels of degradation.

Third, \textbf{Runtime} refers to the time taken in seconds to convert digital files into DNA sequences and subsequently back into their original binary format. The encoding time reflects the algorithm's efficiency in transforming binary data into DNA sequences, while the decoding time evaluates how quickly and accurately the original data can be reconstructed from the encoded DNA.

Fourth and last, \textbf{Cost} pertains to the estimated expenses associated with chemically or enzymatically synthesizing and sequencing DNA. This measure is influenced by factors such as the length of the oligonucleotides and the average costs associated with sequencing and synthesis processes. Currently, storage costs are not included in this estimate.

This mapping ensures that our app produces indicators directly actionable for system architects. For instance, the \textit{information density} metric can be combined with DNA Data Storage Alliance's scalability clause to expose codecs whose density gains are achieved only at super-quadratic computational cost.
Collectively, these evaluation criteria form a comprehensive framework that facilitates a thorough analysis of the strengths and weaknesses of the DNA data storage codecs assessed in this study.

\section{Results}\label{sec-results}
% ==========================================
% All in past tense

% The data results

% The relationship between the results
% ==========================================
We organize our findings into two distinct sections. The first section details the benchmark results of various DNA codecs using default parameters, highlighting their performance metrics and comparative advantages. The second section introduces our app, outlining its key features, functionalities, and inherent limitations.
%This structured approach allows for a comprehensive understanding of both the encoding methodologies and the practical application of our developed tool within the context of DNA data storage encoding methods analysis.

\subsection{Benchmark}\label{sec-benchmark-results}

The benchmark results provided several critical insights into the performance of various DNA codecs integrated within the app. We emphasize that all performance measures examined in this study are strongly influenced by the full set of relevant codec parameters.
Below we systematically present the key findings derived from our analysis. This includes a detailed examination of specific performance metrics, such as performance variability, base error, computational efficiency, and cost efficiency. Additionally, we conduct a comparative analysis of the codecs, highlighting their respective advantages and disadvantages based on empirical data.

%\textcolor{red}{WORK IN PROGRESS from this point onwards until \autoref{sec-tool-results}}

%\textbf{Performance Variability}\quad Significant differences in information densities were observed among the codecs. The observed dispersion aligns with the scalability limits predicted analytically for Levenshtein-distance clustering and LT decoding~\cite{shomorony2022}, reinforcing DNA Data Storage Alliance's call for sub-quadratic codecs in archival contexts. Some methods achieved higher information densities, while others demonstrated lower ratios, indicating a trade-off between data size reduction and computational resource requirements. 

\begin{figure}[htb]
    \centering
    \includegraphics[width=\linewidth]{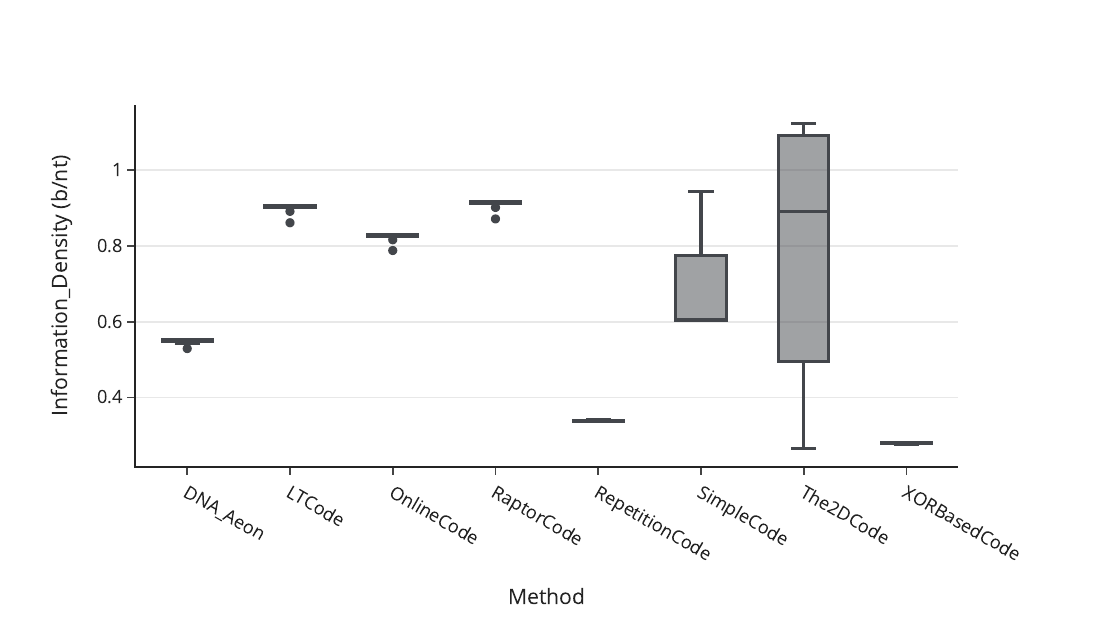}
    \caption{\textbf{Overall information density of the evaluated codecs in the no-error condition.} The plot summarizes the information density achieved by each codec before degradation is introduced, allowing direct comparison across all eight integrated codecs.}
    \label{fig:information-density-overall}
\end{figure}

\paragraph{Performance Variability} The benchmark results demonstrated significant variability in information density across the integrated codecs and benchmark data sets. In the absence of error, defined herein as an error multiplier of 0.0, all eight codecs successfully completed the benchmark and produced measurable information-density values, allowing direct comparison across the entire set of codecs as seen in \autoref{fig:information-density-overall}. NOREC4DNA LT-based and Raptor-based codecs produced consistently high information densities across most files, reaching approximately 0.9~b/nt. In the no-error setting, NOREC4DNA Online-based codec demonstrated a similar pattern, although with slightly lower values. DNA-Aeon produced a lower but stable information density, while Repetition and XOR-based codecs occupied the lower end of the information density range, reflecting the redundancy introduced by these codecs.

\begin{figure}[htb]
    \centering
    \includegraphics[width=\linewidth]{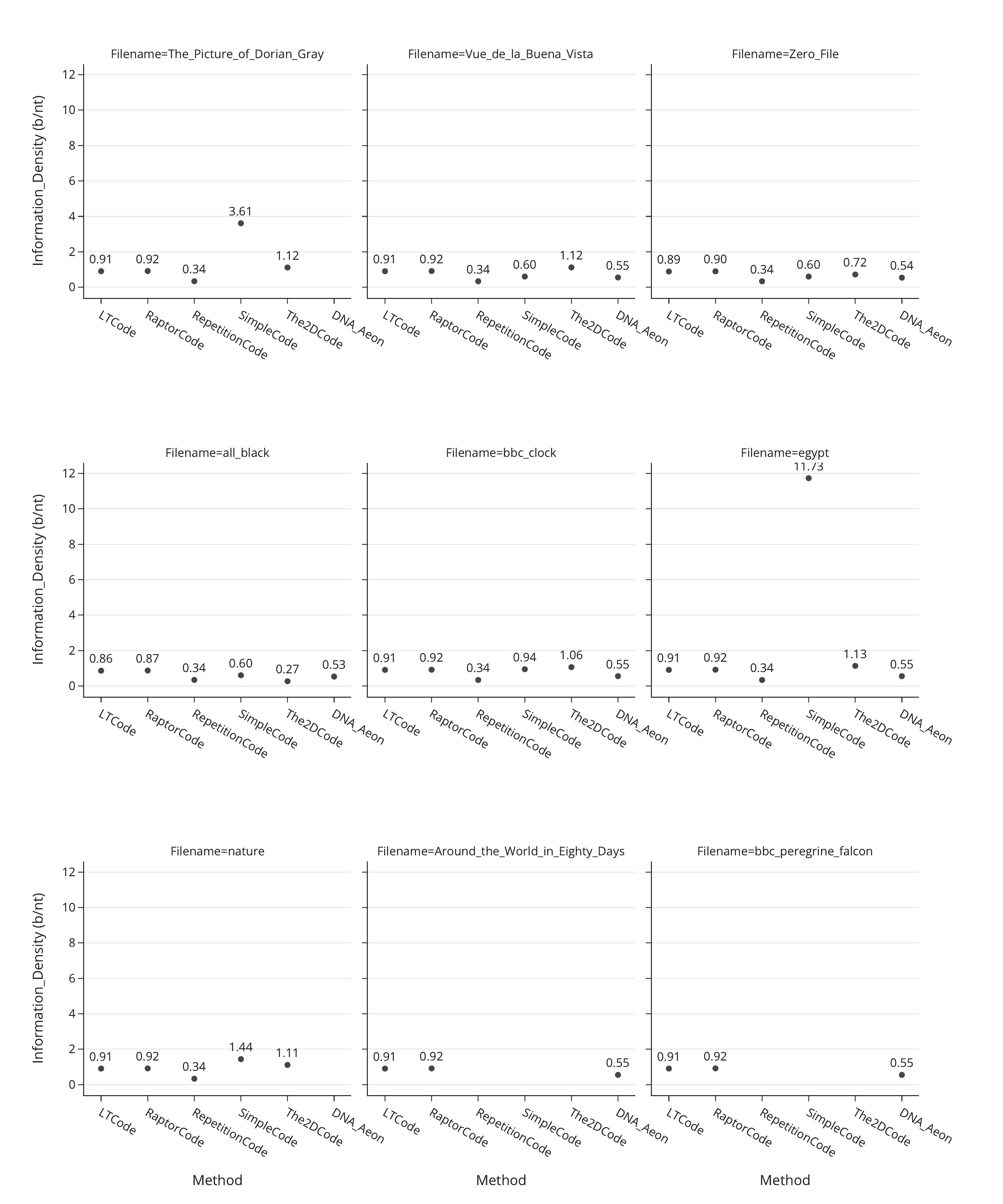}
    \caption{\textbf{File-level information density across codecs at error multiplier 0.5.} The plots show how information density varies across benchmark files and codecs. Missing codec entries indicate runs that did not complete successfully under the default configuration used in this study.}
    \label{fig:information-density-by-file}
\end{figure}

The most significant file-dependent variation was observed for Simple and 2D codecs. In a subset of files, as shown in \autoref{fig:information-density-by-file}, these codecs exhibited significantly higher information-density values compared to the remaining codecs. Conversely, in other files, their information densities were closer to the overall range observed across the benchmark set. This finding suggests that the information density was not solely determined by the choice of codec; rather, it was influenced by the interplay between the coding scheme and the input file's structure. This is consistent with the benchmark design, which intentionally included files with different formats, sizes, byte-value distributions, and entropy profiles.

Across a range of increasing error multipliers, the information density values for codecs that were successfully completed remained largely comparable to those in the no-error condition. This is expected since the information density is computed from the original file size and the length of the encoded DNA sequences, whereas the error multiplier affects the degradation and decoding stages. Therefore, changes observed in error-multiplier charts should be interpreted primarily in relation to the presence or absence of completed codec runs. Missing entries indicate codec--file--error combinations that did not complete successfully under the default configuration we used in this study.

These results show that information density alone is insufficient for evaluating the performance of codecs. NOREC4DNA LT-based and Raptor-based codecs demonstrated the most consistent information density profile across the benchmark set, while Simple and 2D codecs achieved higher information densities in selected cases but with stronger file dependence. DNA-Aeon showed a lower but more stable information density profile. These patterns underscore the need for codec comparison tools that evaluate storage efficiency alongside robustness and runtime, rather than ranking codecs solely by information density.

\begin{figure}[htb]
    \centering
    \includegraphics[width=\linewidth]{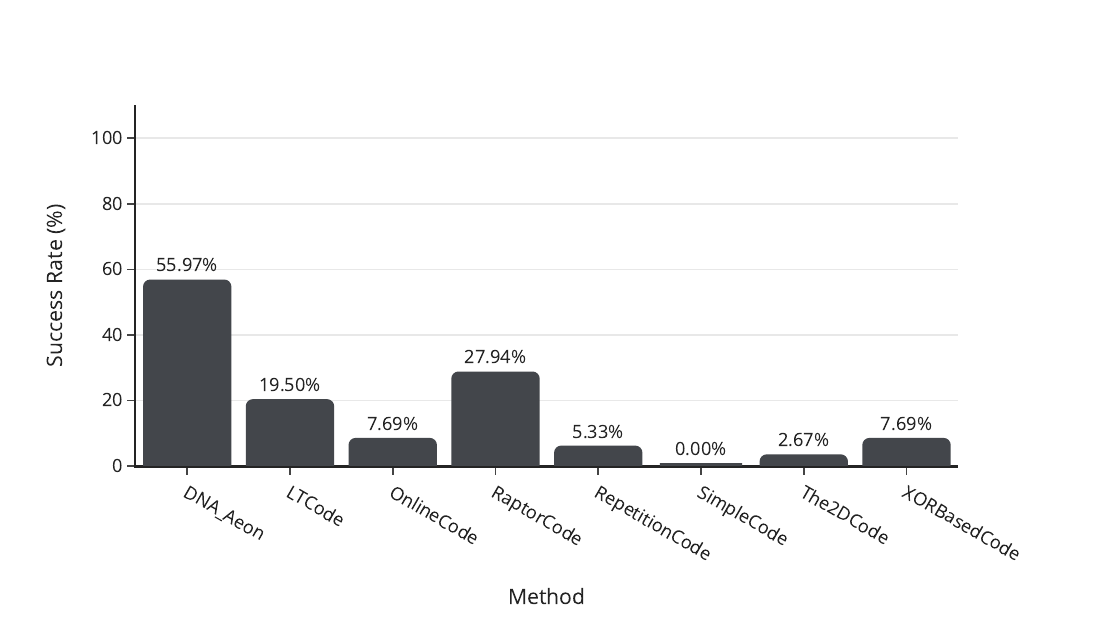}
    \caption{\textbf{Overall success rate across evaluated codecs.} Success rate was calculated by comparing the original binary data with the decoded output after degradation and decoding. DNA-Aeon achieved the highest overall success rate, followed by NOREC4DNA Raptor-based and LT-based codecs.}
    \label{fig:success-rate-overall}
\end{figure}

\paragraph{Base Error and Data Integrity} The results of the success-rate analysis showed that the robustness of the codecs differed significantly across the integrated methods. Unlike information density, which remained relatively stable across completed runs, the success rate showed a high degree of sensitivity to the introduction of base errors. As shown in \autoref{fig:success-rate-overall}, across the comprehensive benchmark, DNA-Aeon achieved the highest overall success rate at 55.97\%, followed by NOREC4DNA Raptor-based codec at 27.94\% and NOREC4DNA LT-based codec at 19.50\%. NOREC4DNA Online-based and XOR-based codecs each achieved 7.69\%, Repetition codec achieved 5.33\%, 2D codec achieved 2.67\%, and Simple codec did not successfully recover any tested file under the evaluated conditions. This approach directly aligns with the success-probability and error-tolerance criteria recommended for DNA codec evaluation~\cite{SNIA2025}.

\begin{figure}[htbp]
    \centering
    \includegraphics[width=\linewidth]{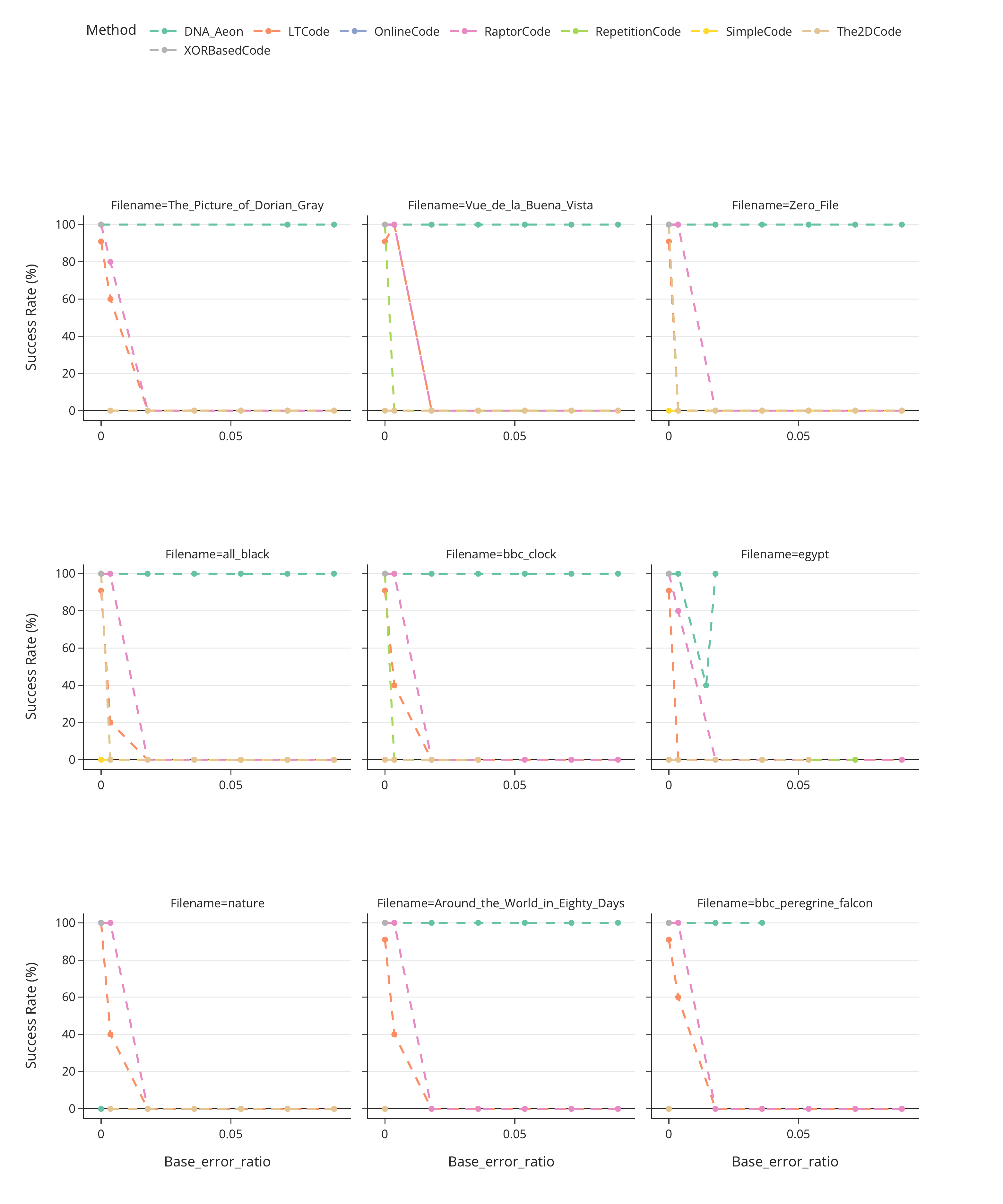}
    \caption{\textbf{Effect of base-error ratio on codec success rate across benchmark files.} The plots show how data recovery changes as base-error ratio increases for each file and codec. Most codecs show a sharp reduction in success rate after errors are introduced, while DNA-Aeon maintains the most consistent recovery profile across the evaluated conditions.}
    \label{fig:base-error-success}
\end{figure}

The base-error analysis further showed that most codecs experienced a sharp decline in success rate once errors were introduced as illustrated in \autoref{fig:base-error-success}. In the absence of errors, several codecs successfully recovered the original data. However, as the base-error ratio increased, successful recovery became increasingly concentrated in a smaller subset of methods. DNA-Aeon showed the most consistent recovery profile across the evaluated base-error range, while NOREC4DNA Raptor-based LT-based codecs demonstrated successful recovery only under more limited conditions. The remaining codecs showed low or inconsistent recovery once degradation was introduced.

These results demonstrate that high information density does not necessarily imply high data integrity. Some codecs that achieved competitive density values showed limited recovery under degraded conditions, while DNA-Aeon achieved the strongest success-rate profile despite its lower information density. Therefore, the benchmark indicates a clear trade-off between storage efficiency and robustness, reinforcing the need to evaluate DNA storage codecs using both density and data-recovery metrics.

%\textbf{Runtime Analysis}\quad The runtime for encoding varied considerably depending on the size of the uploaded files and the number of selected codecs. This variability highlights the importance of selecting appropriate coding strategies based on specific use cases and resource availability.

\paragraph{Computational Efficiency} The runtime results showed substantial differences in computational performance across codecs, files, encoding and decoding stages. Runtime was measured separately for encoding and decoding, and the results are presented on a logarithmic scale because execution times varied by several orders of magnitude. As illustrated in \autoref{fig:runtime-no-error-a}, in the absence of errors, smaller and lower-intricacy files generally required shorter execution times. In contrast, larger or more intricate data types, such as the \texttt{egypt} video and audio files, exhibited significantly extended runtime across several codecs.

The separation between encoding and decoding further showed that computational cost was not evenly distributed across the two stages. Some codecs completed encoding relatively quickly but required substantially longer decoding time, while others showed the opposite pattern. NOREC4DNA LT-based, Online-based, and Raptor-based codecs generally required longer runtimes for larger files, reflecting the additional computational overhead associated with rateless erasure coding and reconstruction. DNA-Aeon also showed increased decoding time under degraded conditions, particularly for larger files and files with more complex byte-value distributions. In contrast, Repetition, Simple, and XOR-based codecs frequently exhibited superior processing speeds upon completion. However, their lower success-rate profile indicates that runtime efficiency alone does not imply practical robustness.

As the error multiplier increased, the runtime results became more uneven and more sparse, as illustrated in \autoref{fig:runtime-no-error-b}.
This pattern reflects both the additional computational burden introduced by degraded sequences and the fact that some codec--file--error combinations did not complete under the default configuration used in this study. Consequently, runtime must be interpreted jointly with completion and success-rate results. A codec that completes quickly but fails to recover the original data offers limited practical value, while a codec with higher runtime may be preferable when it provides stronger data integrity under error.

Overall, the runtime analysis shows that computational efficiency is a major source of variation among DNA storage codecs. The benchmark therefore highlights a three-way trade-off between information density, data integrity, and runtime. NOREC4DNA Raptor-based and LT-based codecs provided comparatively strong information density but required greater computational time for larger files. DNA-Aeon achieved the strongest success-rate profile but could become computationally expensive during decoding under degraded conditions. Simpler codecs were faster in completed runs, but this advantage was offset by weaker recovery performance.

\begin{figure}[t]
    \centering
    \begin{subfigure}[t]{\textwidth}
        \centering
        \includegraphics[width=\linewidth]{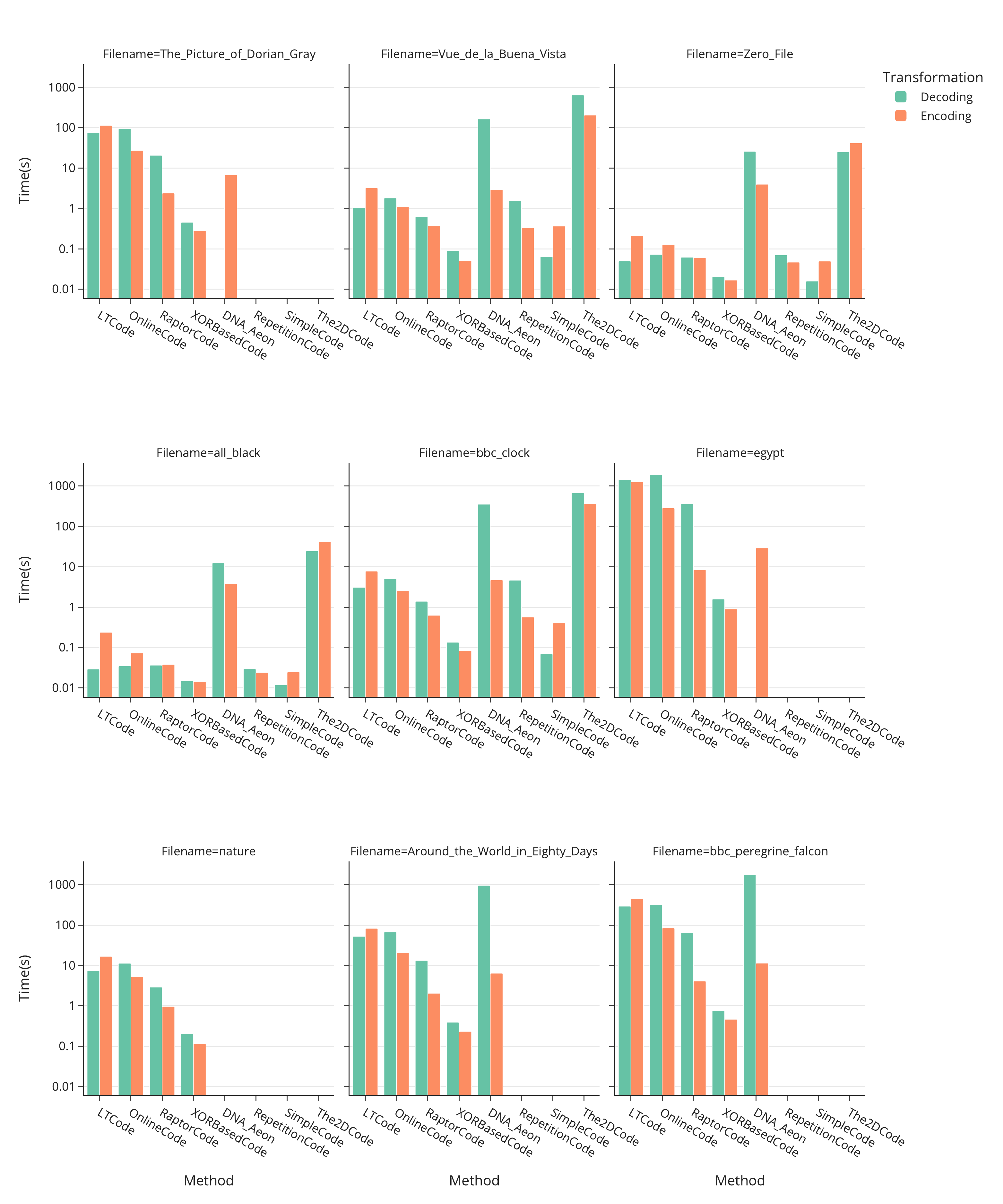}
        \caption{Error multiplier 0.0.}
        \label{fig:runtime-no-error-a}
    \end{subfigure}
\end{figure}
\begin{figure}[t]\ContinuedFloat
    \begin{subfigure}[t]{\textwidth}
        \centering
        \includegraphics[width=\linewidth]{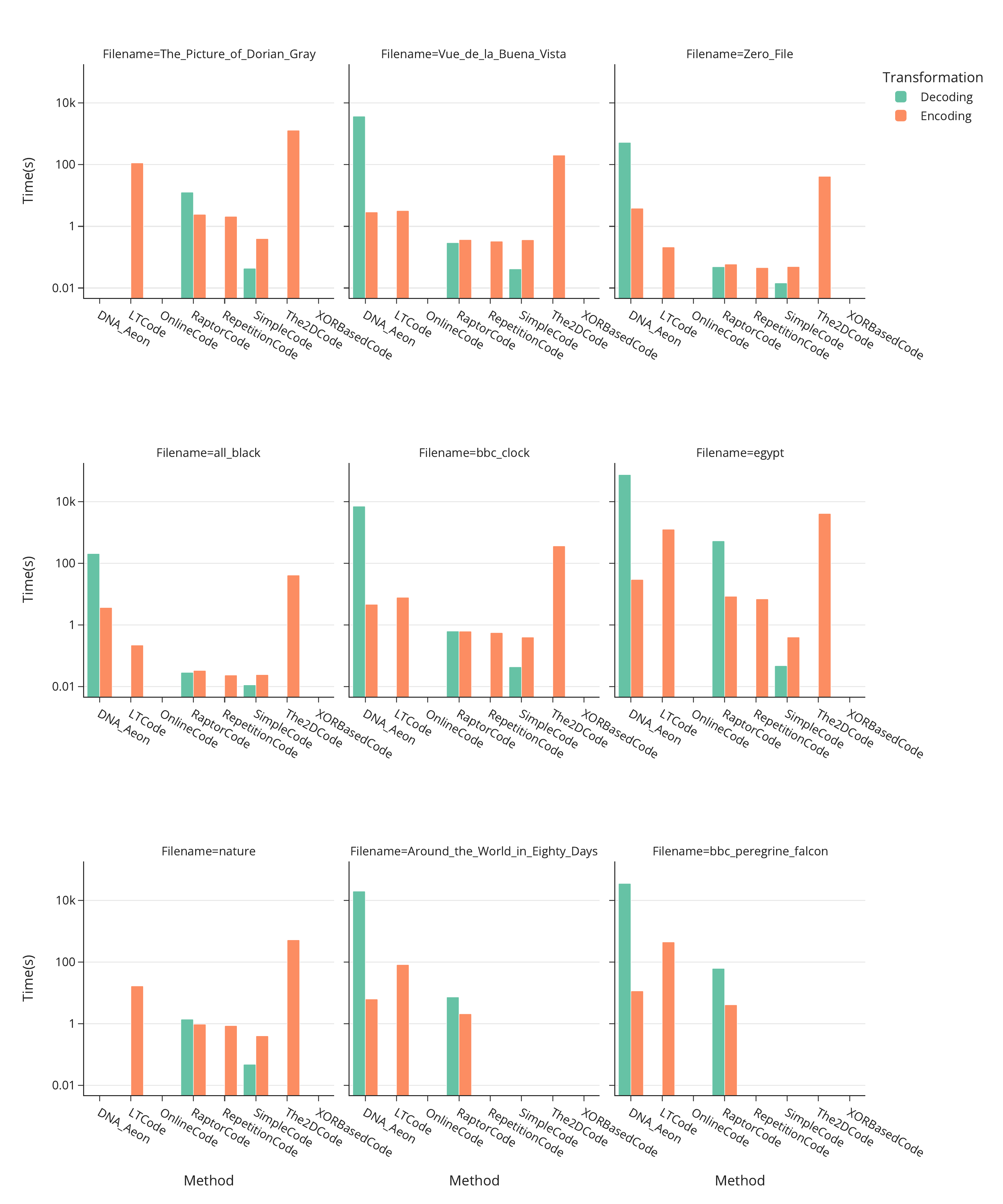}
        \caption{Error multiplier 0.5.}
        \label{fig:runtime-no-error-b}
    \end{subfigure}
    \caption{\textbf{Encoding and decoding runtime across codecs under different error conditions.} Runtime is shown on a logarithmic scale because execution times varied by several orders of magnitude across codecs and benchmark files. \autoref{fig:runtime-no-error-a} shows the no-error condition, while \autoref{fig:runtime-no-error-b} shows the corresponding runtime profile at error multiplier 0.5. Missing entries indicate codec--file--error combinations that did not complete successfully under the default configuration used in this study.}
    \label{fig:runtime-comparison}
\end{figure}

%\textbf{Cost Efficiency}\quad Cost metrics associated with each codec revealed that certain methods were more economical than others. This information is crucial for researchers and organizations aiming to optimize their data storage solutions while managing operational costs effectively.

\paragraph{Cost Efficiency} The cost analysis showed that estimated storage cost varied substantially across codecs, benchmark files, encoding and decoding stages. As with runtime, cost values were presented on a logarithmic scale because the estimates spanned several orders of magnitude. As illustrated in \autoref{fig:cost-no-error-a}, across the benchmark set, larger files exhibited substantially higher estimated costs; for example, the \texttt{egypt} video file and the larger audio and text files were the most expensive. In contrast, smaller or highly uniform files, such as \texttt{all\_black} and \texttt{Zero\_File}, generally required lower estimated costs.

Across codecs, encoding costs were consistently higher than decoding costs. This pattern reflects the fact that the encoded DNA sequence, including payload, headers, and redundancy, directly affects the amount of DNA that must be synthesized. Codecs that introduced more redundancy or generated longer encoded sequences therefore produced higher cost estimates, while more compact sequences generally reduced the cost burden. However, the relationship between cost and information density was not uniform across all files, indicating that cost was also influenced by file size, codec-specific overhead, and the number and length of generated DNA sequences.

The cost results also show that low cost cannot be interpreted independently of data recovery. While some codecs produced comparatively lower costs for specific files, these advantages were not always accompanied by high success rates under degraded conditions. Conversely, DNA-Aeon showed stronger recovery performance but did not always produce the lowest cost estimate. Therefore, cost efficiency must be interpreted alongside information density and success rate, rather than as a standalone measure of codec performance.

As the error multiplier increased, the cost results became more sparse, as illustrated in \autoref{fig:cost-no-error-b}, reflecting the same pattern observed in the information-density and runtime results. Missing entries indicate codec--file--error combinations that did not complete successfully under the default configuration used in this study. Overall, the cost analysis highlights a practical trade-off in DNA data storage: codecs that improve robustness or introduce additional redundancy may increase synthesis-related cost, while lower-cost configurations may provide weaker protection against degradation.

\begin{figure}[t]
    \centering
    \begin{subfigure}[t]{\textwidth}
        \centering
        \includegraphics[width=\linewidth]{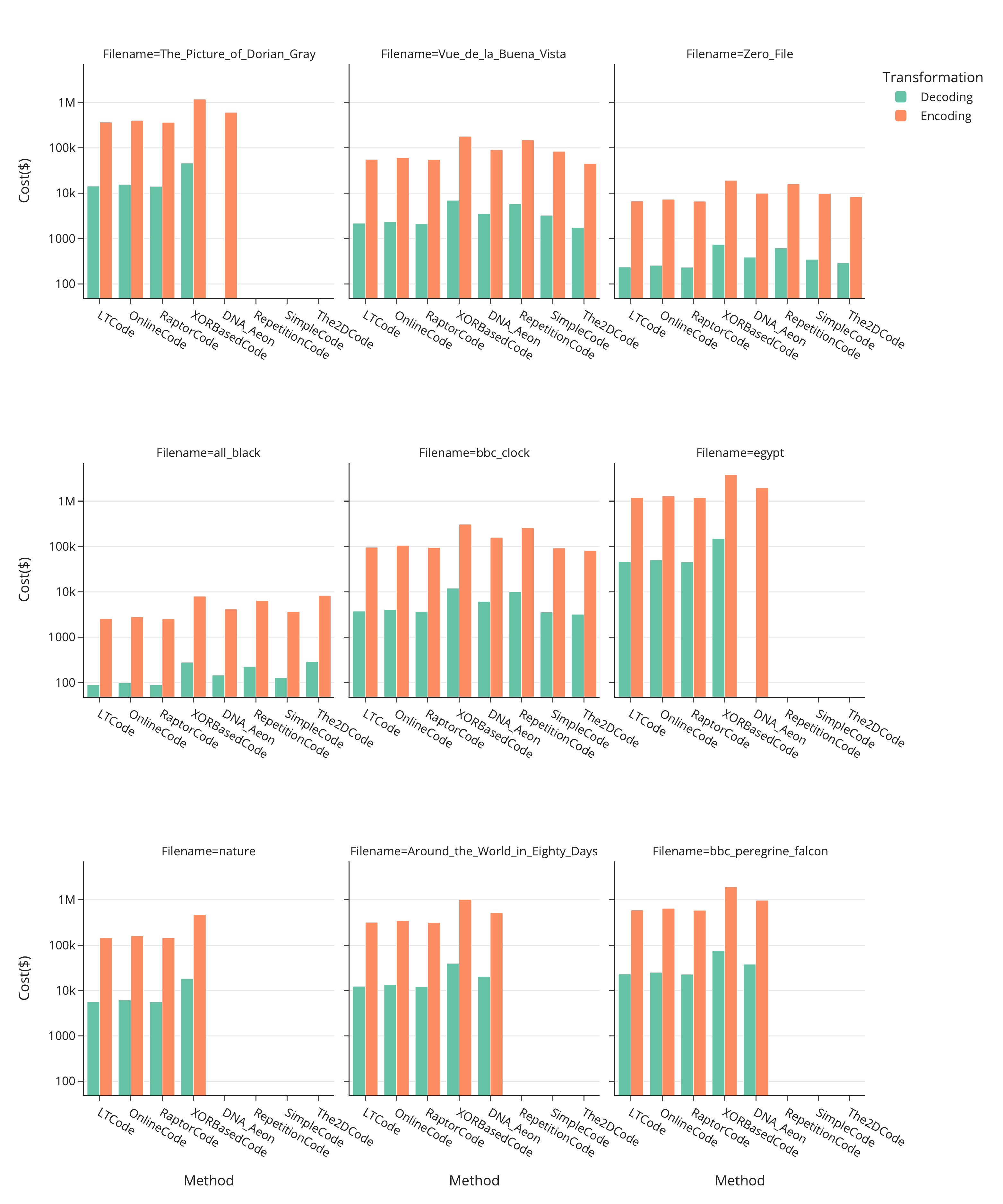}
        \caption{Error multiplier 0.0.}
        \label{fig:cost-no-error-a}
    \end{subfigure}
\end{figure}
\begin{figure}[t]\ContinuedFloat
    \begin{subfigure}[t]{\textwidth}
        \centering
        \includegraphics[width=\linewidth]{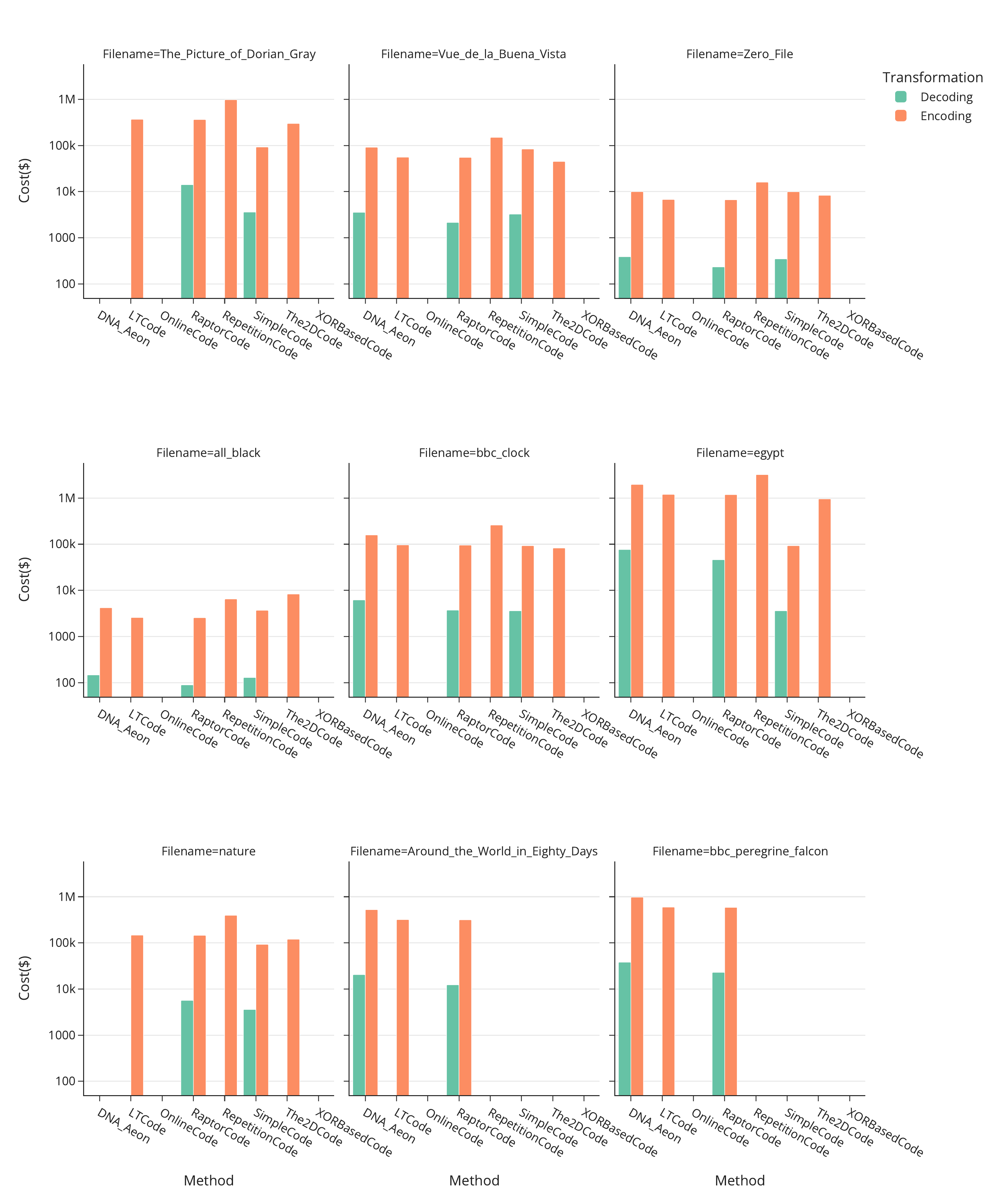}
        \caption{Error multiplier 0.5.}
        \label{fig:cost-no-error-b}
    \end{subfigure}
    \caption{\textbf{Estimated encoding and decoding cost across codecs under different error conditions.} Costs are shown on a logarithmic scale because estimates varied by several orders of magnitude across codecs and benchmark files. \autoref{fig:cost-no-error-a} shows the no-error condition, while \autoref{fig:cost-no-error-b} shows the cost profile at error multiplier 0.5. Missing entries indicate codec--file--error combinations that did not complete successfully under the default configuration used in this study.}
    \label{fig:cost-comparison}
\end{figure}

%\textbf{Impact of Codec Choices}\quad The results underscored how different codec choices could influence not only the efficiency of data storage but also the integrity of the data during retrieval. The ability to compare these metrics directly aids users in making informed decisions when selecting codecs.
\paragraph{Impact of Codec Choices} The combination of these results shows that codec choice has a direct effect on storage efficiency, data integrity, computational performance, and estimated cost. No single codec dominated across all evaluation criteria. NOREC4DNA LT-based and NOREC4DNA Raptor-based codecs achieved consistently high information densities, but their performance must be interpreted alongside their runtime and recovery behaviour. DNA-Aeon achieved the strongest overall success-rate profile, despite producing lower information density than the rateless erasure codes. In contrast, less complex or more compact schemes were sometimes computationally efficient or cost-efficient for specific files. However, these methods showed weaker recovery under degraded conditions.

It is also important to note that some codecs appear in figures summarizing multiple metrics (\textit{e.g.}, encoding/decoding runtime or cost) despite not achieving successful end-to-end reconstruction. For instance, the Simple codec is included because it was able to perform encoding and decoding under certain error regimes and for certain files. However, the decoded outputs never matched the original input files, as evidenced by~\autoref{fig:success-rate-overall}. Consequently, although measurable results are reported, this codec did not achieve a successful encode--decode cycle, and its results should therefore be interpreted with caution. Similar behavior was also observed for other codecs, indicating that this limitation is not unique to a single method.

These findings suggest that codec selection is application-dependent. For use cases where storage density is the primary constraint, NOREC4DNA LT-based and Raptor-based codecs may be preferable. For scenarios where recovery under degradation is more important, DNA-Aeon provides a more robust profile under the evaluated configuration. For rapid testing or low-complexity files, simpler codecs may still be useful, provided that their lower error tolerance is acceptable. Therefore, the app's comparative framework is valuable because it allows users to evaluate codecs across multiple dimensions rather than selecting a method based on a single metric.
Overall, the benchmark demonstrates that codec evaluation requires a multi-criteria approach in which information density, success rate, runtime, and cost are interpreted collectively.

\subsection{Web Application}\label{sec-tool-results}

First, the application's modular nature, abstracted implementations, and use of modern programming languages enable easier adoption by experts in the field. We also utilized Docker, which automates app deployment within portable containers. By doing so, our work encourages reproducibility and collaboration by ensuring consistent behavior across different environments.

Second, even though the app's architecture is designed and finished, improving the app's utility remains a work in progress. This is due to the rapidly evolving field of DNA data storage, which leads to constant advancements in codecs. Our plan is to continuously update the app with new DNA codecs while also ensuring smooth collaborations with other scientists.
To achieve this, we have open-sourced the web app and its data, written extensive documentation, and developed the app to adhere to commonly used Python conventions and standards (\textit{e.g.}, Python Enhancement Proposal~(PEP) 8, PEP 20, PEP 257, PEP 287~\cite{results-python-book}, \textit{etc.}).

Third, we have clearly outlined the process for integrating currently unavailable DNA codecs into our app. By leveraging a version control system (\textit{i.e.}, git) and a GitHub repository, along with a detailed step-by-step guide, we aim to encourage other researchers and practitioners to actively contribute to and enhance the app.

Fourth and last, we developed a Python script that automates the recalculation of the benchmark data set whenever a new codec is successfully integrated into the web app. This script features a user-friendly dashboard that allows users to easily access both previously calculated benchmark results and the results generated by the newly implemented codecs. Users have the option to update the pre-calculated benchmark results with the new data, and seamlessly push these changes to the app's main repository. To further support extensive benchmarking efforts, we also provide an additional script designed to run the pre-calculation on an High-Performance Computing (HPC) Slurm environment~\cite{slurm}, which is particularly beneficial for benchmarks that may require substantial computing time or resources. This automation not only enhances efficiency but also streamlines the integration of new codecs into the existing framework. By simplifying this evaluation process, we aim to encourage frequent updates and enhancements to the app, keeping it at the forefront of DNA data storage technology.

\section{Discussion}\label{sec-discussion}
% ==========================================
% Answer to the hypothesis or research question. It has to be the same as in the introduction and in present tense.

% Discuss how our results support this answer

% Discuss how other's results support this answer

% Discuss the possible limitations and future work
% ==========================================

%DNA data storage is an interdisciplinary field that bridges coding theory, cryptography, informatics, molecular biology, bioinformatics, and computer science. Its potential hinges on the continued advancement of DNA synthesis and sequencing technologies. Although the theoretical foundations of DNA data storage are still developing, the field is rapidly emerging as a promising area of research. Central to this technology are the encoding and decoding methods, which serve as the critical link between digital information and DNA molecules. Over the past decade, these methods have been the primary focus of research, with various approaches offering strengths in information density, technical compatibility, and storage robustness. However, the lack of standardized comparative evaluation systems for these methods has hindered further progress and practical application.

The interdisciplinary nature of DNA data storage requires rigorous evaluation frameworks that bridge the gap between theoretical foundations and practical implementation challenges. Our unified evaluation app fills a critical gap in the field by providing standardized benchmarking capabilities while maintaining the flexibility essential for ongoing methodological innovation.

Furthermore, our platform fosters collaboration among researchers and serves as a guidance or reference app for various storage needs. The study optimizes a series of established DNA data storage codecs at the software engineering level, offering a universal application for DNA data storage. The app ensures compatibility with existing technologies and stability for long-term storage by establishing an evaluation and analysis system based on widely accepted key parameters in the field. The study uses different file types and formats to evaluate classical DNA data storage codecs, enabling systematic analysis and comparison.

However, such evaluations are complicated by the fact that many DNA codecs have different configuration options that can significantly impact performance. This variability poses a challenge for benchmarking because direct comparisons may be confounded by differences in parameter settings. To address this issue, the study carefully selects optimal or default configurations for each codec, prioritizing settings that demonstrate consistent and robust performance across a wide range of benchmark files. This strategy ensures a fair, standardized evaluation framework that remains practically relevant, facilitating meaningful comparisons while preserving the diversity inherent to each codec.
Therefore, future additions to the benchmark could incorporate parameter sweeps, adaptive optimization strategies, or context-dependent tuning (\textit{e.g.}, error profiles, synthesis, and sequencing constraints). This would provide a more comprehensive evaluation of each codec's potential performance under diverse scenarios.

It is important to emphasize that the benchmark results presented in this study, while encompassing multiple codecs and diverse file types, should not be interpreted as definitive indicators of the absolute quality or performance of the individual codecs. Due to the large configuration space and file-dependent behavior of DNA storage methods, we did not exhaustively optimize parameters for every codec--file combination. Instead, we employed a set of well-justified, ``sane'' default configurations for each codec, as described earlier, to ensure consistency and practical feasibility. Consequently, the primary contribution of this work lies not in ranking codecs, but in demonstrating the capabilities of the proposed app as a unifying tool and platform that enables standardized, reproducible, and extensible evaluation across heterogeneous methods, thereby supporting more transparent and comparable benchmarking practices in the field.

This study aimed to solve the challenge of unifying and standardizing methods and their evaluation without stifling innovation. Although standardization is crucial for consistently evaluating different codecs, maintaining flexibility is equally important for exploring new solutions and identifying emerging problems. Maintaining this balance is critical to ensuring the field continues to evolve and adapt to new discoveries.
The structured design of the wrapper functions used in this study supports scalability and ensures that new methods can be incorporated with minimal disruption. As the field of DNA data storage advances, this flexible framework will allow the research community to efficiently explore and implement innovative encoding and decoding strategies, thereby advancing the capabilities of DNA as a data storage medium.

Integrating the DNA Data Storage Alliance's codec taxonomy enabled us to reposition our platform as not only a performance web application, but also a compliance verification framework. By scoring each method against codified community criteria, we create a living baseline that can adapt as consensus standards (\textit{e.g.}, ISO/IEC JTC1 SC31) evolve. This mitigates the risk of future format obsolescence, as discussed in Landsman and Strauss~\cite{Landsman2023}.

The benchmark results further demonstrate why such a standardized framework is necessary. Codec performance was not governed by a single metric: methods that achieved high information density did not always achieve high recovery success, low runtime, or low estimated cost. NOREC4DNA LT-based and Raptor-based codecs showed consistently high information-density profiles, while DNA-Aeon achieved the strongest overall success-rate performance under the evaluated degradation settings. These differences indicate that codec choice is inherently application-dependent and should be guided by the intended balance between storage efficiency, data integrity, computational feasibility, and cost. Thus, the main contribution of the app is not to identify a universally best codec, but to make these trade-offs visible under a common evaluation framework.

%Additionally, the performance measures employed in this study were essential for providing a comprehensive evaluation of the DNA data storage encoding and decoding methods. These measures included encoding and decoding time, file sizes, synthesis cost, and data integrity. Encoding and decoding time are important for determining the viability of the methods in real-world scenarios where processing speed is a significant factor. File sizes generated during the encoding process offered a measure of data information density, which is particularly important when managing large data sets. Understanding synthesis cost is vital for evaluating the economic feasibility of DNA data storage methods, particularly for large-scale or long-term data storage solutions. Finally, high data integrity is essential, as it ensures that the stored data can be reliably recovered without errors, a fundamental requirement for any storage technology. Together, these measures provided a robust framework for evaluating the performance of various DNA data storage encoding methods.

However, while comprehensive within its scope, our evaluation framework faces several inherent limitations that warrant consideration.
First, although reliance on MESA's default parameters for degradation simulation is justified by the need for methodological consistency, such an approach may not accurately reflect the full diversity of storage conditions encountered across various deployment scenarios. In future work, alternative solutions, such as DNA Storalator~\cite{dna-storalator}, will be explored, with the goal of providing the end user with additional DNA storage simulation options and yielding a more comprehensive picture of the DNA storage landscape.

Second, the comparison relies on selected default or representative codec configurations. This was necessary because the parameter space of several codecs is too large for exhaustive exploration, but it also means that some failed or missing runs may have succeeded under alternative settings. In particular, some codec--file--error combinations did not complete under the evaluated configuration, and several implementations returned limited diagnostic information when failures occurred. Future work should therefore include more systematic parameter sweeps and more explicit failure classification, distinguishing between decoding failure, timeout, unsupported error types, file-size limitations, and implementation-level errors.

Third, the binary data representation through NumPy arrays, chosen for compatibility across codecs, introduces computational overhead that obscures the true performance characteristics of optimized implementations.

Fourth and finally, although we attempted to implement as many DNA codecs as possible, we were unable to include several others that remain highly relevant to the field. Some noteworthy examples are Gungnir~\cite{Gungnir-codec} and DNAformer~\cite{new-method-daniella-2025}, among others. The former uses proof-of-work idea to address substitution, insertion, and deletion errors in a DNA sequence, while the latter combines deep neural networks, specifically a transformer model with error-correcting codes, to significantly enhance information retrieval from DNA-stored data. Our ongoing efforts aim to integrate more such methods into our framework to further strengthen the comprehensiveness and utility of our work.

\section{Conclusion}\label{sec-conclusion}
% ==========================================
%Conclusions may be used to restate your hypothesis or research question, restate your major findings, explain the relevance and the added value of your work, highlight any limitations of your study, describe future directions for research and recommendations. 

% How the study helps solve the problem from the introduction. What is the value of this work

% What is the next step of this research
% ==========================================
The development of the proposed unified evaluation app is a foundational step toward standardizing the assessment of DNA data storage while preserving the methodological flexibility essential for continued innovation. Our platform implements the five-dimensional evaluation framework proposed by the DNA Data Storage Alliance, providing the research community with a reproducible benchmarking system that directly addresses real-world deployment considerations.

Our analysis of eight codecs using default parameters and across various data types shows that no single algorithm can optimize all performance dimensions simultaneously. This finding underscores the importance of selecting application-specific methods and highlights the value of comparative evaluation frameworks. The observed performance patterns, particularly regarding scalability limitations and cost-efficiency trade-offs, provide actionable insights for algorithm developers and system architects alike.

The app's modular architecture and standardized wrapper functions facilitate the integration of emerging codecs while maintaining evaluation consistency. This design philosophy enables the platform to adapt to technological advances while preserving historical benchmarking data for longitudinal analysis. Aligning with the DNA Data Storage Alliance criteria positions our framework to support the field's transition from experimental demonstrations to commercial implementations.

Looking ahead, successfully deploying DNA storage technologies critically depends on standardized evaluation methodologies that can guide research priorities and industrial adoption decisions. Our contribution provides the community with a foundation for evidence-based method selection and a platform for the collaborative advancement of the field. As synthesis costs decline and sequencing throughput increases, standardized evaluation frameworks will be essential for realizing DNA storage's transformative potential.

The open-source nature of our implementation encourages community participation and ensures that the platform evolves alongside technological developments. By establishing this evaluation infrastructure, we enable the systematic assessment necessary to transform DNA data storage from a promising laboratory technique into a reliable, standardized technology for long-term information preservation.

Future research should expand the evaluation framework to incorporate emerging synthesis technologies and novel error-correction schemes, while maintaining backward compatibility with existing benchmarking data. Additionally, integrating machine learning techniques optimized for graphics processing unit (GPU)-accelerated encoding and decoding would enhance the platform's performance and adaptability. This would enable more efficient and accurate analysis of increasingly complex DNA data storage methods.

\backmatter

\bmhead{Code availability}

Source code, help, and documentation are available at \url{https://github.com/AAnzel/UNACORM/}. The repository also contains the source code necessary to reproduce all examples shown in this work. It is licensed under the GNU General Public License, Version 3.0, and can be manipulated, improved, and extended freely by any user.

\bmhead{Availability of data and materials}

The data used in this study can be found at \url{https://github.com/AAnzel/UNACORM/tree/master/Data/Original}.

%\bmhead{Supplementary information}
%If your article has accompanying supplementary file/s, please state so here. 
%Please refer to Journal-level guidance for any specific requirements.

\bmhead{Authors' contributions}
C.A. and A.A. wrote the manuscript, designed the app, and evaluated the results. C.A., A.A., and I.T.H. conducted the experiments and validated the findings. I.T.H. and A.A. developed the main app. C.A., A.A., and G.H. designed the user experience. L.W., L.S., D.S., K.E., S.B., F.K., R.S., M.W., and M.S. reimplemented or adapted individual codecs that are part of the main app. C.A. and A.A. supervised the reimplementation. D.H. and G.H. supervised the project. C.A, A.A, M.W, M.S., K.E., R.S., L.W., D.H., B.F., and G.H. proofread, and revised the manuscript. All authors read and approved the final manuscript.
%\bmhead{Acknowledgments}

\bmhead{Funding}
This work was partially financially supported by the LOEWE program of the State of Hesse (Germany) in the MOSLA research cluster.

\bmhead{Conflict of interest}
The authors declare that they have no known competing financial interests or personal relationships that could have appeared to influence the work reported in this paper.

%\begin{appendices}

%\section{Section title of first appendix}\label{secA1}

%An appendix contains supplementary information that is not an essential part of the text itself but which may be helpful in providing a more comprehensive understanding of the research problem or it is information that is too cumbersome to be included in the body of the paper.




%\end{appendices}


\bibliography{sn-bibliography}% common bib file


\end{document}